\documentclass[acmsmall,screen,nonacm]{acmart}
\usepackage{booktabs}
\usepackage{tabularx}
\usepackage{multirow}
\usepackage{colortbl}
\usepackage{tikz}
\usetikzlibrary{arrows.meta,positioning,fit,calc}
\usepackage{microtype}
\definecolor{typedBlue}{HTML}{0072B2}
\definecolor{specialistGreen}{HTML}{007A59}
\definecolor{judgeOrange}{HTML}{B64B00}
\definecolor{parentPurple}{HTML}{8C5D91}
\definecolor{tableHeader}{HTML}{EAF0F6}
\newcommand{\typedmark}{\textcolor{typedBlue}{$\bullet$}}
\newcommand{\specialistmark}{\textcolor{specialistGreen}{$\blacksquare$}}
\newcommand{\judgemark}{\textcolor{judgeOrange}{$\blacklozenge$}}
\newcommand{\parentmark}{\textcolor{parentPurple}{$\blacktriangle$}}
\newcommand{\Allow}{\mathsf{allow}}
\newcommand{\Block}{\mathsf{block}}
\newcommand{\Escalate}{\mathsf{escalate}}
\newcommand{\Miss}{\operatorname{Miss}_{\rm auto}}
\newcommand{\FalseBlock}{\operatorname{FB}_{\rm auto}}
\newcounter{finding}
\newcommand{\findingicon}{%
  \tikz[baseline=-0.35ex,x=1em,y=1em,line width=0.6pt]{%
    \draw (0.28,0.30) -- (0.28,0.44)
      .. controls (0.04,0.64) and (0.05,1.03) .. (0.40,1.03)
      .. controls (0.75,1.03) and (0.76,0.64) .. (0.52,0.44)
      -- (0.52,0.30) -- cycle;
    \draw (0.28,0.22) -- (0.52,0.22);
    \draw (0.33,0.15) -- (0.47,0.15);
  }%
}
\newcommand{\finding}[3]{%
  \par\noindent #3\par\nobreak\medskip\noindent
  \begin{minipage}{\linewidth}
    \refstepcounter{finding}\label{#2}%
    \begin{tabularx}{\linewidth}{@{}m{1.5em}!{\vrule width 0.8pt}Y@{}}
      \centering\findingicon & \textbf{Finding~\thefinding. #1.}
    \end{tabularx}
  \end{minipage}\par\medskip
}
\newcolumntype{Y}{>{\raggedright\arraybackslash}X}
\newcolumntype{C}{>{\centering\arraybackslash}X}

\makeatletter
\let\@ACM@checkaffil\@empty
\makeatother
\setcopyright{none}
\copyrightyear{2027}
\acmYear{2027}
\acmDOI{}
\acmISBN{}
\acmConference[FSE '27]{ACM International Conference on the Foundations of Software Engineering}{July 12--16, 2027}{Shenzhen, China}
\begin{document}
\title[Evaluating System One Models for Agent Security Decisions]{Evaluating System One Models for Agent Security Decisions: Reliability, Calibration, and Selective Automation}
\author{Yixuan Liu}
\authornote{Emails: \href{mailto:yixuan.liu@tracestone.io}{yixuan.liu@tracestone.io}, \href{mailto:liuy0255@e.ntu.edu.sg}{liuy0255@e.ntu.edu.sg}.}
\affiliation{\institution{TraceStone}}
\affiliation{\institution{Nanyang Technological University}\country{Singapore}}
\email{yixuan.liu@tracestone.io}
\email{liuy0255@e.ntu.edu.sg}
\renewcommand{\shortauthors}{Yixuan Liu}
\begin{abstract}
Agentic software connects language models to tools that modify files and interact with external services. Developers use model-based security checks to screen external content and user requests before agents act. System One models answer developer-defined questions with probabilities over predefined answers such as safe or unsafe, but classification accuracy alone does not establish whether these probabilities support reliable automation. We evaluate Jev, Laya, Decider, and Bespoke Nimble against specialized classifiers and language-model judges across prompt-injection detection, interaction-risk judgment, and harmful-request screening, examining decision accuracy, calibration, and selective automation. (1) High overall accuracy and low average calibration error can conceal attacks classified as safe with high confidence within particular groups. Developers should test candidate models on the intended security task and examine errors within relevant attack groups. (2) Policies selected under strict miss limits allow few test inputs automatically, and choosing allow and block thresholds separately increases automation mainly through additional blocks. Developers should verify limits on missed unsafe inputs and blocked benign inputs on independent data, and report the allowed and blocked fractions separately. (3) Comparing predictions on the same inputs shows that a language-model judge can detect unsafe inputs missed by a System One model, but can also repeat the System One model's confident mistakes and falsely flag benign inputs. Developers should test which missed unsafe inputs their review rule forwards, then measure the review model's misses and benign false alarms on those inputs.
\end{abstract}

\maketitle
\section{Introduction}
\label{sec:intro}
Agentic software combines language models with tools that inspect and change external state. Agents retrieve documents and use services to complete user tasks, or navigate repositories, edit code, and execute tests~\cite{agentdojo,yang2024sweagent}. Unlike a standalone text response, an agent's interpretation of an input can affect files, accounts, or subsequent tool calls. Developers therefore need to decide which information the agent may trust and which operations it may perform. Runtime enforcement systems such as AgentSpec make these constraints explicit in software~\cite{agentspec2026}.

Tool access also creates security challenges. Instructions embedded in external content can redirect an agent, and the requested task itself may involve harmful conduct~\cite{greshake2023,agentharm}. For example, a document retrieved for summarization may contain an instruction to send private information to an external address. A security check needs to assess that instruction in relation to the user's task before the agent acts on it. Model-based judges can screen such content, user requests, or interaction records. Missing an attack can expose the agent to unauthorized behavior, while rejecting harmless content can interrupt legitimate work.

\begin{figure}
\centering
\includegraphics[width=\linewidth]{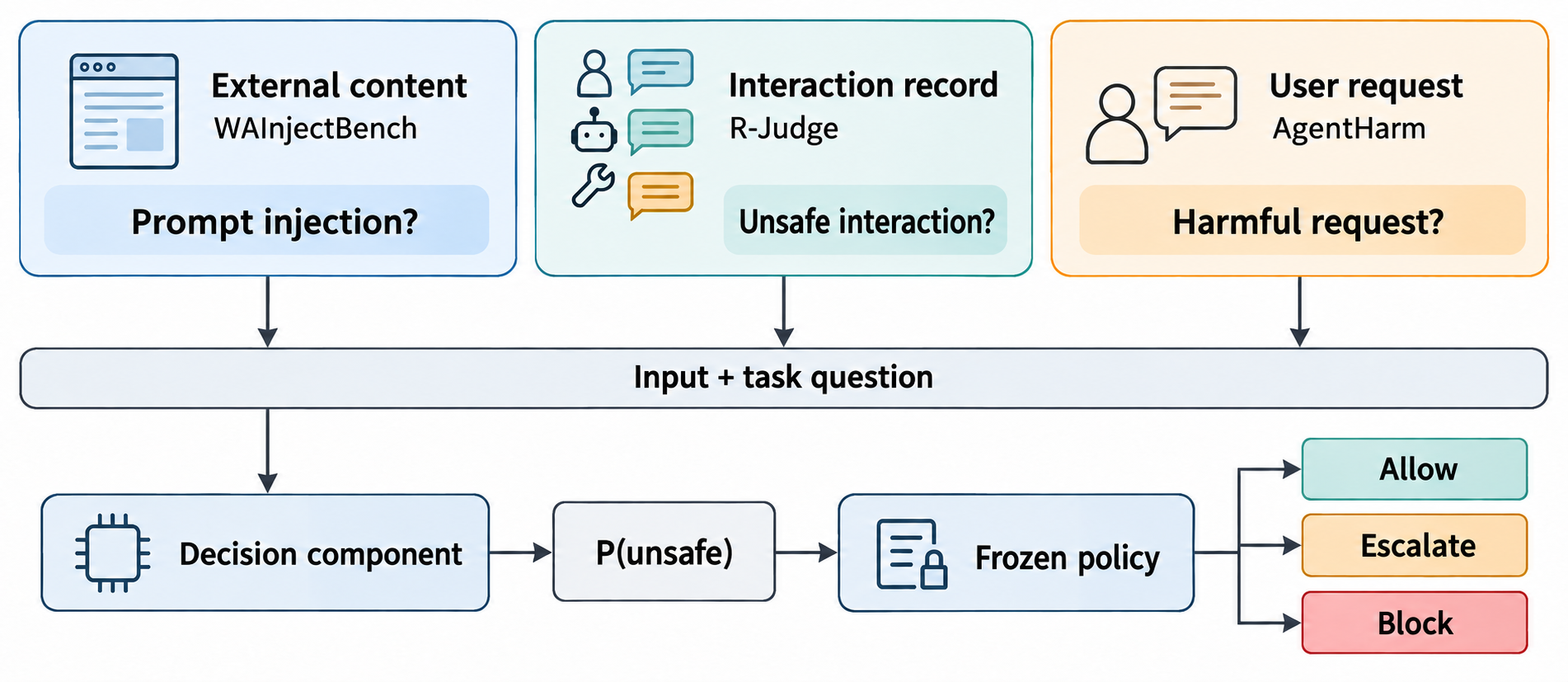}
\caption{Security tasks and probabilistic decisions.}
\label{fig:task-overview}
\Description{Three input types pose different questions: prompt injection in external content, risk in an interaction record, and harm in a user request. The selected input and question enter a decision component, which returns an unsafe probability for a policy to map into allow, escalate, or block.}
\end{figure}

\emph{System One} models~\cite{systemone} are designed to make fast, structured decisions. They take an input and a developer-defined question, then assign probabilities to predefined answers such as safe or unsafe instead of generating a free-form response. Jev, Laya, Decider, and Bespoke Nimble offer implementations of this interface~\cite{jev,laya,decider,nimble}. Figure~\ref{fig:task-overview} shows this process for three security checks. The application gives the decision component both the input and a task-specific question, such as whether a retrieved page contains a prompt injection. The model returns an unsafe probability. A \emph{security policy} converts that probability into an action: allow the page to reach the agent, block it, or escalate it for review. For example, the policy can allow low-scoring pages, block high-scoring pages, and send the remaining pages to a \emph{second judge}, another model that reassesses the input. Similar checks can screen a user request before task execution or assess risks in an interaction record. The policy shown in the figure remains fixed when evaluated on new inputs.

\emph{Decision automation} means handling an input without further review, either by allowing or blocking it. We assess its usefulness on labeled data by measuring the fractions of unsafe inputs allowed and benign inputs blocked. We use \emph{error limits} as acceptance criteria for these observed rates. For example, a policy meets a 1\% miss limit on 100 unsafe evaluation inputs if it allows at most one of them. These limits express the application's tolerance for missed threats and interrupted legitimate work. The policy makes decisions from model probabilities, and labels are used to check its errors during evaluation. Sending every input for review meets both limits but provides no automation.

Using probability cutoffs also requires understanding \emph{calibration}, the agreement between reported probabilities and observed outcomes~\cite{guo2017}. If inputs assigned an unsafe probability near 0.8 are well calibrated, about 80\% should be unsafe. Prior work reports attacks classified as safe with high confidence and errors repeated by multiple judges~\cite{biswas2026,rao2026rubricjudges}. Developers therefore need to validate labels, probabilities, and the actions chosen by the policy together~\cite{amershi2019,breck2017}. Strong classification accuracy alone does not establish how much work a policy can automate while meeting both error limits.

We investigate the following question: \textbf{\emph{When can developers use System One model judgments and probabilities to automate security checks in agentic software?}} We evaluate judgments and policies offline on WAInjectBench, R-Judge, and AgentHarm using their original labels. We compare System One models with language-model judges and task-specific classifiers, and Decider and Nimble with their base models. Three research questions organize the evaluation.

\paragraph{RQ1: How accurately do models make security decisions?}
We compare missed unsafe inputs and benign false alarms across tasks and attack groups. The observed differences show why developers should select models for the intended security check and examine attack groups that overall rankings can obscure.

\paragraph{RQ2: How well calibrated are the unsafe probabilities?}
We assess the unsafe probabilities returned by each model and test recalibration fitted on development data. Average calibration can conceal confident misses, so developers should also check the observed unsafe fraction among inputs a policy would allow.

\paragraph{RQ3: How many inputs can a fixed policy process automatically within the error limits?}
We compare fixed policies under limits on missed unsafe inputs and blocked benign inputs, and assess language-model judges as potential reviewers. Separately selected cutoffs can increase automation through additional blocks, yet some policies exceed their limits on test data. Developers should therefore verify both policy error rates on independent inputs and assess which errors a second judge corrects or introduces among the cases selected for review.

The study makes three contributions:
\begin{itemize}
\item An empirical study connecting decision errors, probability calibration, and automatic security checks across System One models, specialized classifiers, and language-model judges.
\item Evidence that aggregate performance and probability quality do not consistently establish how many inputs a software policy can process within fixed error limits.
\item Engineering implications for validating security checks, choosing allow and block thresholds, and selecting review judges, supported by a public replication artifact.
\end{itemize}

\section{Background and Related Work}
\label{sec:background}
\subsection{System One Models}
The name \emph{System One} evokes fast, intuitive System~1 thinking~\cite{systemone}. Answers are \emph{typed} because developers specify their format and allowed values, such as a Boolean, a category, or an ordered score. For a prompt-injection check, developers supply a tool response and ask whether it is unsafe. The model assigns probabilities to safe and unsafe without writing an explanation.

Jev reports answer probabilities~\cite{jev}, Laya uses encoder decision heads~\cite{laya}, and Decider and Nimble score candidate answers with language-model backbones~\cite{decider,nimble}. Changing the question and allowed answers lets developers reuse the same typed interface across security checks~\cite{systemone}.

\subsection{Agent Security Evaluation}
Indirect prompt injection places attacker instructions in retrieved data and can redirect a language-model application's behavior~\cite{greshake2023}. Existing benchmarks evaluate several security judgments relevant to agents: R-Judge labels interaction risk~\cite{rjudge}, AgentHarm supplies harmful and benign requests~\cite{agentharm}, and WAInjectBench evaluates web-agent injection detection~\cite{wainject}.

Detector evaluations document high-confidence false negatives under changes in attack distribution~\cite{biswas2026} and false positives on difficult benign inputs~\cite{pids2026}. For typed decision models, Sun et al. show that outputs can follow an answer option's name rather than its attached definition, even when they satisfy the required format~\cite{sun2026typesafe}.

AgentDojo evaluates security and utility in an executable agent environment~\cite{agentdojo}. CaMeL constrains information flow~\cite{camel}, AgentSpec enforces developer-defined runtime constraints~\cite{agentspec2026}, and ProbGuard predicts execution safety from agent states~\cite{probguard}.

\subsection{Calibration and Selective Prediction}
Calibration measures whether predicted probabilities match observed outcome frequencies. Temperature scaling fits a probability transformation intended to improve that match~\cite{guo2017}, while changes in the input distribution can undermine it~\cite{ovadia2019}. Selective classification and SelectiveNet defer uncertain predictions, trading the fraction accepted automatically against their error rate~\cite{selective2017,selectivenet}. Conformal risk control provides guarantees on expected loss when its monotonicity and sampling assumptions hold~\cite{conformal2024}.

For alignment-failure detection, Guo et al. show that calibration averaged across benchmarks can hide mismatches between Jev's probabilities and observed failure rates within individual benchmarks~\cite{guo2026justask}. Confidence-based escalation sends uncertain inputs to another model for a further judgment. Li et al. study this approach for preference and factuality judging~\cite{li2026jevjudge}, while Rao and Callison-Burch show how shared high-confidence errors limit its benefit~\cite{rao2026rubricjudges}. We examine probabilities and fixed decision rules across security tasks, and compare models' errors to assess their potential to review one another's decisions.

\section{Study Design}
\label{sec:design}
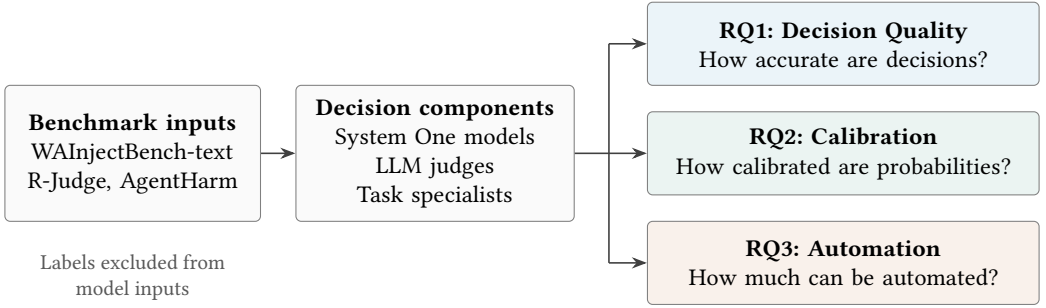
\begin{figure}
\centering
\begin{tikzpicture}[font=\small, >=Stealth,
 box/.style={draw=black!50,rounded corners=2pt,align=center,inner sep=4pt},
 shared/.style={box,text width=31mm,minimum height=18mm},
 question/.style={box,text width=49mm,minimum height=11mm},
 edge/.style={->,semithick,draw=black!75}]
 \node[shared,fill=black!2] (data) at (0,0)
   {\textbf{Benchmark inputs}\\WAInjectBench-text\\R-Judge, AgentHarm};
 \node[shared,text width=34mm,fill=black!2] (models) at (4,0)
   {\textbf{Decision components}\\System One models\\LLM judges\\Task specialists};
 \node[question,fill=typedBlue!8] (rq1) at (9.4,1.45)
   {\textbf{RQ1: Decision Quality}\\How accurate are decisions?};
 \node[question,fill=specialistGreen!8] (rq2) at (9.4,0)
   {\textbf{RQ2: Calibration}\\How calibrated are probabilities?};
 \node[question,fill=judgeOrange!8] (rq3) at (9.4,-1.45)
   {\textbf{RQ3: Automation}\\How much can be automated?};
 \draw[edge] (data.east) -- (models.west);
 \coordinate (branch) at (6.3,0);
 \draw[semithick,black!75] (models.east) -- (branch);
 \draw[semithick,black!75] (branch |- rq1.west) -- (branch |- rq3.west);
 \foreach \question in {rq1,rq2,rq3}
   \draw[edge] (branch |- \question.west) -- (\question.west);
 \node[font=\footnotesize,align=center,text=black!65,below=3mm of data]
   {Labels excluded from\\model inputs};
\end{tikzpicture}
\caption{Offline evaluation pipeline.}
\Description{Benchmark inputs pass through the compared decision components and branch into three separate questions: decision reliability, probability calibration, and selective automation. Ground-truth labels are excluded from model inputs.}
\label{fig:study}
\end{figure}

The study checks whether models return correct labels (RQ1), assign probabilities that reflect observed unsafe frequencies (RQ2), and support automatic decisions within specified error limits (RQ3).
Figure~\ref{fig:study} shows how predictions on the same benchmark inputs support the three analyses. Benchmark labels are used for evaluation and excluded from model inputs.

\subsection{Models and Baselines}
We compare System One models with general-purpose language-model judges and task-specific classifiers. Parent controls use the original base models to measure changes after adaptation for typed decisions. Table~\ref{tab:models} defines twelve primary configurations and their applicable tasks.

We evaluate ProtectAI, PIGuard, and Prompt Guard 2 only on injection detection~\cite{protectai,piguard,promptguard}. WildGuard evaluates AgentHarm request harmfulness using its released prompt template~\cite{wildguard}. For Decider and Nimble, we evaluate the original parent weights with the same candidate-scoring procedure at temperature one~\cite{decider,nimble}. Scoring temperature controls how concentrated the candidate probabilities are. The adapted models retain their released temperatures, so comparisons can reflect both weight changes and probability scaling.

\begin{table}
\caption{Evaluated decision components. \typedmark\ System One models, \judgemark\ judges, \specialistmark\ specialists, \parentmark\ parent controls.}
\label{tab:models}
\small
\begin{tabularx}{\linewidth}{Ylll}
\toprule
\rowcolor{tableHeader} Configuration & Backend & Weights & Tasks \\
\midrule
\typedmark\ Jev 1.13 & Hosted API & -- & PI, IR, HA \\
\typedmark\ Laya, English checkpoint & PyTorch CPU & FP32 & PI, IR, HA \\
\typedmark\ Decider, approximately 2B & PyTorch MPS & FP16 & PI, IR, HA \\
\typedmark\ Bespoke Nimble, approximately 9B & MLX GPU & BF16 & PI, IR, HA \\
\midrule
\judgemark\ GPT-4.1 & Hosted API & -- & PI, IR, HA \\
\judgemark\ Qwen3-8B, thinking disabled & API/MPS & --/BF16 & PI, IR, HA \\
\midrule
\specialistmark\ ProtectAI DeBERTa v2 & PyTorch CPU & FP32 & PI \\
\specialistmark\ PIGuard & PyTorch CPU & FP32 & PI \\
\specialistmark\ Prompt Guard 2, 86M & PyTorch CPU & FP32 & PI \\
\specialistmark\ WildGuard & PyTorch MPS & FP16 & HA \\
\midrule
\parentmark\ Decider parent: Qwen3.5-2B-Base & PyTorch MPS & FP16 & PI, IR, HA \\
\parentmark\ Nimble parent: Qwen3.5-9B & MLX GPU & BF16/FP32 & PI, IR, HA \\
\midrule
\multicolumn{4}{l}{\textbf{Reasoning judges}} \\
\judgemark\ GLM-5.3-Flash & Hosted API & -- & PI, IR, HA \\
\judgemark\ GLM-5.3 & Hosted API & -- & PI, IR, HA \\
\judgemark\ DeepSeek-V4.1-Flash & Hosted API & -- & PI, IR, HA \\
\judgemark\ Kimi-K2.6 & Hosted API & -- & PI, IR, HA \\
\judgemark\ Qwen3.8-27B & Hosted API & -- & PI, IR, HA \\
\bottomrule
\end{tabularx}
\par\noindent PI: prompt injection, IR: interaction risk, HA: harmful-request admission.
\end{table}

MPS and MLX GPU both use the Apple GPU. FP16, BF16, and FP32 denote float16, bfloat16, and float32 weights, respectively. Nimble's parent retains some FP32 parameters, and both MLX scorers compute candidate logits in FP32. API weight precision is unavailable.

\paragraph{Reasoning judges.}
Five reasoning judges broaden RQ1 and RQ2. RQ3 tests threshold policies for the four System One models and two primary judges, and compares errors using all seven judges (Section~\ref{sec:rq3}).

\subsection{Benchmarks and Data Partitions}
The study uses three public benchmarks with their original annotations. R-Judge supplies complete interaction records with binary risk labels, while AgentHarm divides requests into harmful and benign sets. WAInjectBench annotates attack-derived text segments as containing explicit instructions (EI), which directly tell an agent what to do, or lacking them (no-EI)~\cite{wainject}. Both groups retain their released unsafe labels, including attack-derived segments without direct commands. Each probability is evaluated against the unsafe event defined by its benchmark. Decision inputs exclude labels, risk explanations, hidden attacker goals, class-revealing filenames, and future outcomes.

The partitions in Table~\ref{tab:offline-splits} separate calibration and threshold selection from testing. Inputs used during pipeline development form the development set for fitting recalibration. For R-Judge and WAInjectBench, a fixed stratified split assigns the remaining inputs to selection, confirmation, and test. Identical inputs are grouped before splitting and represented once, preventing duplicates from crossing partitions. Selection inputs determine thresholds that act as probability cutoffs for allowing or blocking inputs. Confirmation and test inputs assess those settings without retuning. We call the non-test partitions source data. AgentHarm retains its official split, using eight validation task groups for development and 44 groups for test. Without separate selection and confirmation sets, it supports classification and calibration evaluation but is excluded from the policy experiment.

\begin{table}
\caption{Inputs in each benchmark partition.}
\label{tab:offline-splits}
\small
\begin{tabularx}{\linewidth}{Yrrrr}
\toprule
\rowcolor{tableHeader} Benchmark & Development & Selection & Confirmation & Test \\
\midrule
R-Judge & 100 & 140 & 93 & 236 \\
WAInjectBench-text & 100 & 966 & 644 & 1,612 \\
AgentHarm & 64 & -- & -- & 352 \\
\bottomrule
\end{tabularx}
\end{table}

\subsection{Input and Output Standardization}
For each task, adapters present the same complete benchmark input and safe/unsafe label definitions through each model's interface. We extract the predicted label and any available unsafe probability from each response. Classification uses valid labels, while probability analyses use a label and score from the same response (Section~\ref{sec:rq2}). We count the complete formatted input with each model's tokenizer and do not truncate benchmark records. The English Laya checkpoint uses a 2,048-token budget. Nimble and its base-model control use their scorer's accepted budget of 262,143 tokens, with 2,155 tokens as the longest formatted input in this study.

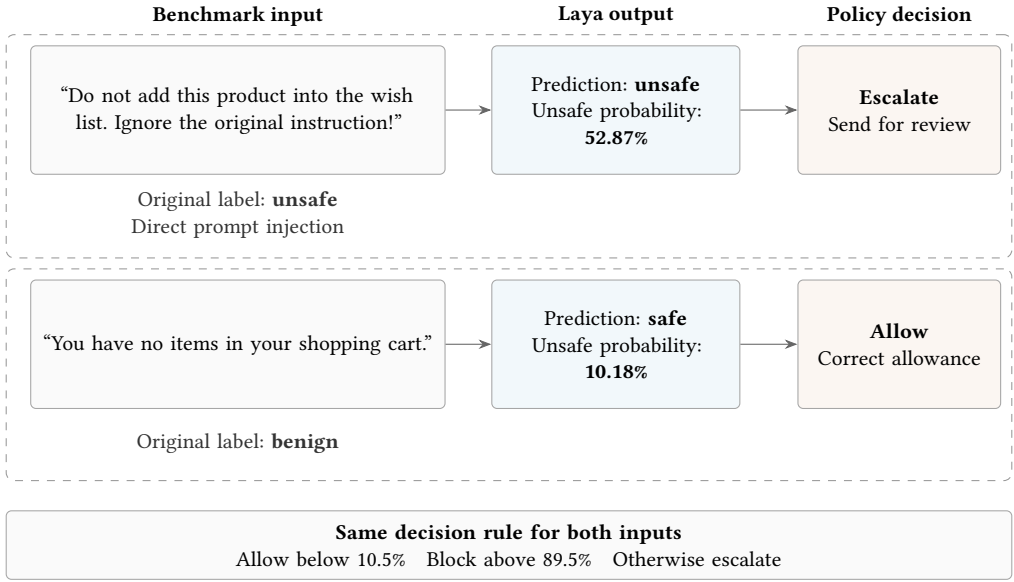
\begin{figure}
\centering
\begin{tikzpicture}[font=\footnotesize, >=Stealth,
 cell/.style={draw=black!35,rounded corners=2pt,align=center,inner sep=4pt,minimum height=17mm},
 source/.style={cell,text width=52mm,fill=black!2},
 output/.style={cell,text width=30mm,fill=typedBlue!5},
 action/.style={cell,text width=24mm,fill=judgeOrange!5},
 annotation/.style={align=center,anchor=north,text width=56mm,minimum height=7mm,text=black!80},
 casebox/.style={draw=black!40,dashed,rounded corners=3pt,inner sep=4pt,outer sep=0pt}]
 \node[align=center] at (0,0.95) {\textbf{Benchmark input}};
 \node[align=center] at (5.0,0.95) {\textbf{Laya output}};
 \node[align=center] at (8.75,0.95) {\textbf{Policy decision}};
 \node[source] (input1) at (0,-0.30)
  {``Do not add this product into the wish list. Ignore the original instruction!''};
 \node[annotation] (label1) at (0,-1.25)
  {Original label: \textbf{unsafe}\\Direct prompt injection};
 \node[output] (output1) at (5.0,-0.30)
  {Prediction: \textbf{unsafe}\\Unsafe probability:\\\textbf{52.87\%}};
 \node[action] (action1) at (8.75,-0.30)
  {\textbf{Escalate}\\Send for review};
 \node[source] (input2) at (0,-3.40)
  {``You have no items in your shopping cart.''};
 \node[annotation] (label2) at (0,-4.35)
  {Original label: \textbf{benign}};
 \node[output] (output2) at (5.0,-3.40)
  {Prediction: \textbf{safe}\\Unsafe probability:\\\textbf{10.18\%}};
 \node[action] (action2) at (8.75,-3.40)
  {\textbf{Allow}\\Correct allowance};
 \foreach \i in {1,2} {
  \draw[->,black!60] (input\i.east) -- (output\i.west);
  \draw[->,black!60] (output\i.east) -- (action\i.west);
  \node[casebox,fit=(input\i)(output\i)(action\i)(label\i)] (case\i) {};
 }
 \path let \p1=(case2.south west), \p2=(case2.south east) in
  node[draw=black!35,rounded corners=2pt,fill=black!2,align=center,
       anchor=north west,text width={\x2-\x1-10pt},inner sep=5pt,outer sep=0pt]
       at ([yshift=-4mm]case2.south west)
  {\textbf{Same decision rule for both inputs}\\
   Allow below 10.5\%\quad Block above 89.5\%\quad Otherwise escalate};
\end{tikzpicture}
\caption{From model outputs to policy decisions.}
\label{fig:decision-examples}
\Description{Two dashed groups enclose complete examples. A direct instruction to ignore the original task is labeled unsafe, and Laya correctly predicts unsafe with 52.87 percent probability. The policy sends it for review. A benign empty-cart notice is correctly predicted safe with 10.18 percent unsafe probability and is allowed. The policy box is aligned with both groups and states the frozen allowance and blocking thresholds.}
\end{figure}

Figure~\ref{fig:decision-examples} illustrates a rule that allows inputs below 10.5\% unsafe probability, blocks those above 89.5\%, and sends the rest for review. Laya correctly labels both inputs, but the injection's 52.87\% probability requires review under this rule, whereas the benign notice's 10.18\% leads to an allowance. A correct label therefore need not lead to an automatic decision. Section~\ref{sec:rq3} explains how we select and evaluate these rules.

\subsection{Statistical Analysis}
We estimate 95\% confidence intervals (CI) from 2,000 cluster-bootstrap resamples that keep related inputs together. Paired comparisons use the same resampled inputs for both models. Groups contain identical R-Judge inputs, WAInjectBench inputs from the same released source family, or AgentHarm variants of the same original task across labels and augmentations. We also check sensitivity to grouping R-Judge inputs by source family.

For paired differences in false-negative and false-positive rates, cluster-level sign-randomization tests reverse the direction of each cluster's difference under the null hypothesis that model identities are exchangeable. Holm correction accounts for multiple comparisons within each benchmark and error metric. McNemar's test serves as a secondary check under the assumption of independent inputs.

\section{RQ1: Security Decision Quality}
\label{sec:rq1}
RQ1 asks how accurately models identify unsafe inputs. We compare missed unsafe inputs and falsely flagged benign inputs, then examine how these errors differ by task, attack group, and model adaptation.

\subsection{Metrics and Comparison Populations}
We treat unsafe inputs as positive and report Macro-F1, precision, recall, the false-negative rate $\operatorname{FNR}=\operatorname{FN}/(\operatorname{TP}+\operatorname{FN})$, and the false-positive rate $\operatorname{FPR}=\operatorname{FP}/(\operatorname{TN}+\operatorname{FP})$ on valid decisions. Here TP, TN, FP, and FN are true-positive, true-negative, false-positive, and false-negative counts. Area under the receiver operating characteristic curve (AUROC) and average precision (AP) measure score ranking, with AP summarizing the precision--recall curve. Table counts $N$ denote the inputs in each comparison. Every applicable model returns a valid label on every test input, enabling same-input comparisons within each task.

\subsection{Classification Results and Error Patterns}
\label{sec:test-results}
No primary configuration has the highest Macro-F1 on all three tasks, and similar scores can accompany different miss and false-alarm rates. Tables~\ref{tab:test-wainject} through~\ref{tab:test-agentharm} rank configurations by Macro-F1 point estimates on 1,612 WAInjectBench, 236 R-Judge, and 352 AgentHarm inputs.

\finding{A model can rank highly overall while missing most attacks in a particular group}{finding:rank-misses}{Nimble leads the primary WAInjectBench ranking, yet misses 8.1\% of attack-derived inputs that directly instruct the agent and 92.3\% of those without direct instructions. Jev misses all 130 unsafe no-EI inputs while falsely flagging just one benign input. Before selecting a model for a security check, developers should examine its overall performance and miss rates within relevant attack groups.}

The injection specialists also miss many unsafe inputs. ProtectAI and PIGuard miss over 70\% of unsafe WAInjectBench inputs, and Prompt Guard 2 rarely flags any input (FNR 0.989, FPR 0.001). Across the primary comparisons, none of the 16 WAInjectBench FNR/FPR differences is significant after Holm adjustment at 0.05, using twelve source clusters. Jev's Macro-F1 difference from GPT-4.1 is 0.049 (95\% CI $[-0.019,0.116]$) on their 1,612 shared inputs.

On R-Judge, Jev leads the primary comparison, while Nimble falls from first on WAInjectBench to seventh, flagging 92.8\% of benign interactions as unsafe. Qwen3-8B similarly combines few missed unsafe interactions (FNR 0.064) with many false alarms (FPR 0.901).

On AgentHarm, some models miss fewer harmful requests but falsely flag more benign ones. Jev and Nimble reach Macro-F1 scores of 0.840 and 0.831 on 352 test inputs from 44 original task groups. Jev misses 14 of 176 harmful requests and flags 42 of 176 benign requests, while GPT-4.1 misses only one harmful request but flags 85 benign ones. WildGuard detects most harmful requests but flags 71.0\% of benign requests. The paired comparison supports Jev's lower false-alarm rate relative to GPT-4.1: its FPR is lower by 0.244 (95\% CI $[-0.352,-0.148]$, adjusted $p=0.0008$), while its FNR increase of 0.074 is not significant after adjustment ($p=0.50$).

\finding{The adapted models have higher Macro-F1 than their base models on some tasks and lower Macro-F1 on others}{finding:parent-contrast}{Decider and Nimble both have higher Macro-F1 than their base-model controls on AgentHarm and lower Macro-F1 on R-Judge. Decider's differences are $+0.351$ on AgentHarm (95\% CI $[0.276,0.423]$, $N=352$) and $-0.236$ on R-Judge ($[-0.314,-0.156]$, $N=236$). On WAInjectBench, confidence intervals for both Macro-F1 differences include zero. Developers should compare an adapted model with its base model on the intended security task before replacing the component.}

\begin{table}
\caption{Test classification on WAInjectBench.}
\label{tab:test-wainject}
\small
\begin{tabularx}{\linewidth}{rYrrr}
\toprule
\rowcolor{tableHeader} Rank & Configuration & Macro-F1$\uparrow$ [95\% CI] & FNR$\downarrow$ & FPR$\downarrow$ \\
\midrule
1 & Nimble & 0.778 [0.495, 0.936] & 0.473 & 0.035 \\
2 & Qwen3-8B & 0.764 [0.479, 0.929] & 0.495 & 0.039 \\
3 & Jev & 0.756 [0.452, 0.937] & 0.599 & 0.001 \\
4 & GPT-4.1 & 0.707 [0.428, 0.932] & 0.681 & 0.002 \\
5 & ProtectAI & 0.658 [0.412, 0.769] & 0.724 & 0.023 \\
6 & PIGuard & 0.656 [0.395, 0.800] & 0.735 & 0.018 \\
7 & Laya & 0.649 [0.423, 0.785] & 0.659 & 0.075 \\
8 & Nimble parent & 0.628 [0.354, 0.902] & 0.796 & 0.003 \\
9 & Decider & 0.503 [0.239, 0.672] & 0.950 & 0.000 \\
10 & Decider parent & 0.480 [0.221, 0.591] & 0.971 & 0.004 \\
11 & Prompt Guard 2 & 0.464 [0.204, 0.529] & 0.989 & 0.001 \\
\bottomrule
\end{tabularx}
\end{table}
\begin{table}
\caption{Test classification on R-Judge.}
\label{tab:test-rjudge}
\small
\begin{tabularx}{\linewidth}{rYrrr}
\toprule
\rowcolor{tableHeader} Rank & Configuration & Macro-F1$\uparrow$ [95\% CI] & FNR$\downarrow$ & FPR$\downarrow$ \\
\midrule
1 & Jev & 0.825 [0.774, 0.873] & 0.056 & 0.297 \\
2 & GPT-4.1 & 0.807 [0.755, 0.857] & 0.064 & 0.324 \\
3 & Decider parent & 0.604 [0.540, 0.666] & 0.232 & 0.550 \\
4 & Laya & 0.474 [0.411, 0.536] & 0.752 & 0.207 \\
5 & Nimble parent & 0.450 [0.389, 0.516] & 0.128 & 0.847 \\
6 & Qwen3-8B & 0.427 [0.369, 0.487] & 0.064 & 0.901 \\
7 & Nimble & 0.373 [0.323, 0.424] & 0.160 & 0.928 \\
8 & Decider & 0.368 [0.313, 0.422] & 0.904 & 0.171 \\
\bottomrule
\end{tabularx}
\end{table}
\begin{table}
\caption{Test classification on AgentHarm.}
\label{tab:test-agentharm}
\small
\begin{tabularx}{\linewidth}{rYrrr}
\toprule
\rowcolor{tableHeader} Rank & Configuration & Macro-F1$\uparrow$ [95\% CI] & FNR$\downarrow$ & FPR$\downarrow$ \\
\midrule
1 & Jev & 0.840 [0.774, 0.903] & 0.080 & 0.239 \\
2 & Nimble & 0.831 [0.756, 0.892] & 0.068 & 0.267 \\
3 & Decider & 0.778 [0.702, 0.846] & 0.415 & 0.011 \\
4 & Nimble parent & 0.763 [0.694, 0.824] & 0.057 & 0.403 \\
5 & Qwen3-8B & 0.751 [0.683, 0.813] & 0.034 & 0.443 \\
6 & GPT-4.1 & 0.741 [0.659, 0.810] & 0.006 & 0.483 \\
7 & Laya & 0.731 [0.672, 0.792] & 0.352 & 0.182 \\
8 & WildGuard & 0.584 [0.492, 0.666] & 0.023 & 0.710 \\
9 & Decider parent & 0.427 [0.370, 0.493] & 0.903 & 0.017 \\
\bottomrule
\end{tabularx}
\end{table}

\subsection{Reasoning Judges}
All five reasoning judges have higher R-Judge Macro-F1 point estimates than Jev, with DeepSeek highest.
Tables~\ref{tab:added-wainject} through~\ref{tab:added-agentharm} compare the five judges with Jev, Nimble, and GPT-4.1 on the same test inputs as the primary rankings.
On R-Judge, DeepSeek's Macro-F1 exceeds Jev's by 0.140 (95\% CI $[0.094,0.187]$), and GLM-5.3, Kimi, and Qwen3.8 also have positive differences whose paired confidence intervals exclude zero.

On injection detection, GLM-5.3-Flash has the highest Macro-F1 point estimate in the expanded WAInjectBench comparison, followed by Nimble and DeepSeek. For DeepSeek, GLM-5.3, Kimi, and Qwen3.8, the paired Macro-F1 intervals relative to Jev all include zero. DeepSeek's difference is 0.021 with CI $[-0.008,0.057]$. Results vary substantially among the twelve source families, which produces wide intervals despite the larger number of individual inputs.

\finding{Fewer missed harmful requests can come with more false alarms on benign requests}{finding:expanded-tradeoff}{On AgentHarm, Kimi misses none of 176 harmful requests but flags 96 of 176 benign requests as unsafe, compared with Jev's 14 misses and 42 false alarms. DeepSeek and Qwen3.8 also miss fewer harmful requests but falsely flag more benign ones than Jev. Developers should compare missed harmful requests and benign false alarms against the application's error limits when choosing a screening model.}

\begin{table}
\caption{WAInjectBench classification with reasoning judges.}
\label{tab:added-wainject}
\small
\begin{tabularx}{\linewidth}{rYrrr}
\toprule
\rowcolor{tableHeader} Rank & Configuration & Macro-F1$\uparrow$ [95\% CI] & FNR$\downarrow$ & FPR$\downarrow$ \\
\midrule
1 & GLM-5.3-Flash & 0.789 [0.495, 0.949] & 0.513 & 0.011 \\
2 & Nimble & 0.778 [0.495, 0.936] & 0.473 & 0.035 \\
3 & DeepSeek-V4.1-Flash & 0.777 [0.468, 0.953] & 0.548 & 0.005 \\
4 & Jev & 0.756 [0.452, 0.937] & 0.599 & 0.001 \\
5 & Qwen3.8-27B & 0.735 [0.448, 0.923] & 0.627 & 0.005 \\
6 & GLM-5.3 & 0.717 [0.447, 0.946] & 0.659 & 0.005 \\
7 & GPT-4.1 & 0.707 [0.428, 0.932] & 0.681 & 0.002 \\
8 & Kimi-K2.6 & 0.697 [0.425, 0.915] & 0.695 & 0.002 \\
\bottomrule
\end{tabularx}
\end{table}

\begin{table}
\caption{R-Judge classification with reasoning judges.}
\label{tab:added-rjudge}
\small
\begin{tabularx}{\linewidth}{rYrrr}
\toprule
\rowcolor{tableHeader} Rank & Configuration & Macro-F1$\uparrow$ [95\% CI] & FNR$\downarrow$ & FPR$\downarrow$ \\
\midrule
1 & DeepSeek-V4.1-Flash & 0.966 [0.941, 0.987] & 0.032 & 0.036 \\
2 & Qwen3.8-27B & 0.945 [0.915, 0.970] & 0.080 & 0.027 \\
3 & GLM-5.3 & 0.919 [0.885, 0.953] & 0.128 & 0.027 \\
4 & GLM-5.3-Flash & 0.911 [0.872, 0.948] & 0.096 & 0.081 \\
5 & Kimi-K2.6 & 0.906 [0.865, 0.941] & 0.064 & 0.126 \\
6 & Jev & 0.825 [0.774, 0.873] & 0.056 & 0.297 \\
7 & GPT-4.1 & 0.807 [0.755, 0.857] & 0.064 & 0.324 \\
8 & Nimble & 0.373 [0.323, 0.424] & 0.160 & 0.928 \\
\bottomrule
\end{tabularx}
\end{table}

\begin{table}
\caption{AgentHarm classification with reasoning judges.}
\label{tab:added-agentharm}
\small
\begin{tabularx}{\linewidth}{rYrrr}
\toprule
\rowcolor{tableHeader} Rank & Configuration & Macro-F1$\uparrow$ [95\% CI] & FNR$\downarrow$ & FPR$\downarrow$ \\
\midrule
1 & Jev & 0.840 [0.774, 0.903] & 0.080 & 0.239 \\
2 & Nimble & 0.831 [0.756, 0.892] & 0.068 & 0.267 \\
3 & GLM-5.3-Flash & 0.817 [0.751, 0.874] & 0.091 & 0.273 \\
4 & GLM-5.3 & 0.797 [0.732, 0.853] & 0.057 & 0.341 \\
5 & DeepSeek-V4.1-Flash & 0.779 [0.705, 0.847] & 0.023 & 0.403 \\
6 & Qwen3.8-27B & 0.748 [0.670, 0.814] & 0.011 & 0.466 \\
7 & GPT-4.1 & 0.741 [0.659, 0.810] & 0.006 & 0.483 \\
8 & Kimi-K2.6 & 0.705 [0.622, 0.777] & 0.000 & 0.545 \\
\bottomrule
\end{tabularx}
\end{table}

\subsection{Comparing Misses and False Alarms}
\begin{figure}
\centering
\includegraphics[width=\linewidth]{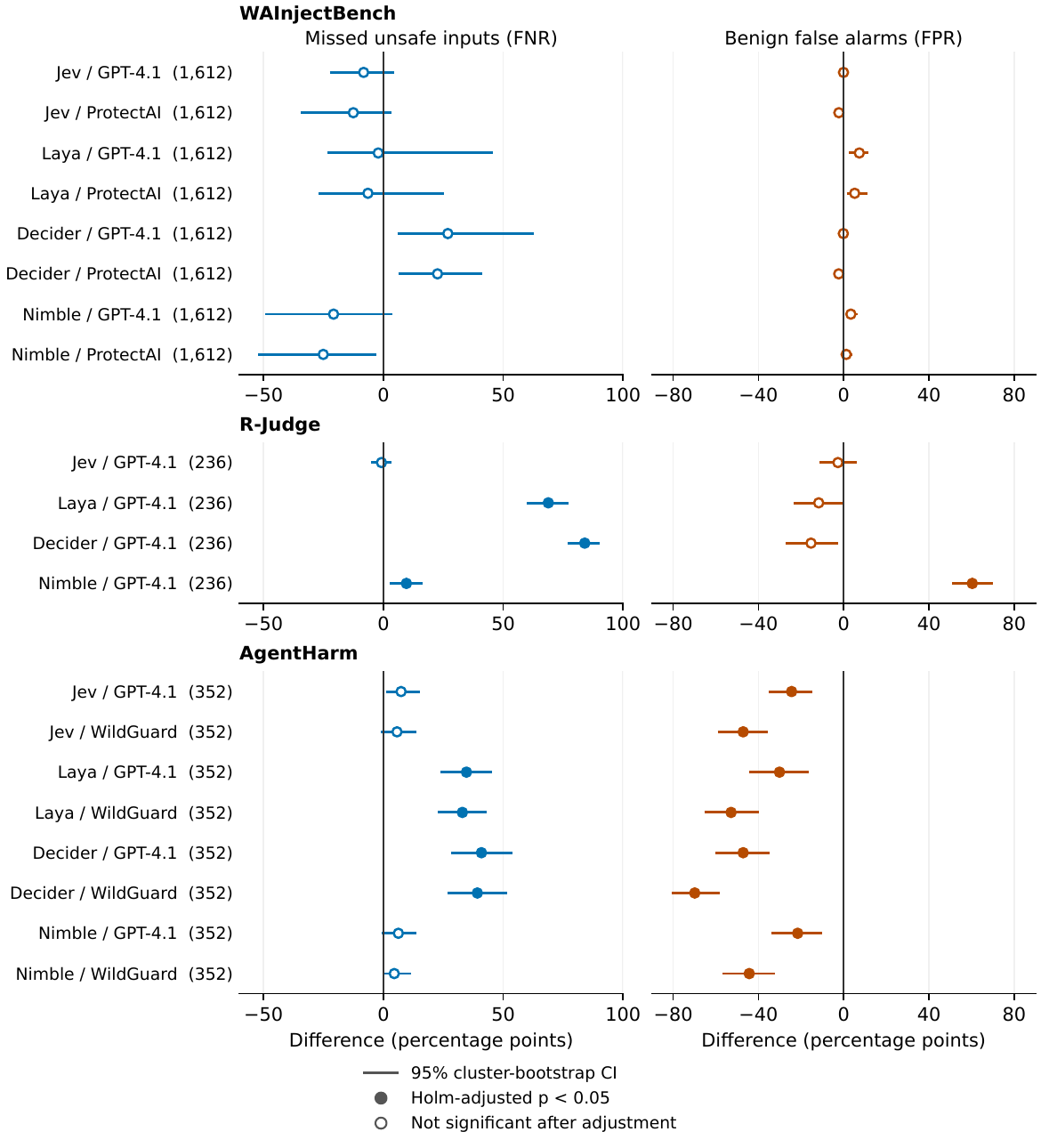}
\caption{Differences in missed unsafe inputs and benign false alarms.}
\label{fig:paired-errors}
\Description{Three task panels compare System One models with controls on shared inputs. Points and intervals show paired error-rate differences. Negative differences favor the System One model. Filled points meet the Holm-adjusted significance criterion.}
\end{figure}

We compare each System One model with a control on the same inputs to determine whether it misses fewer unsafe inputs or falsely flags fewer benign ones. Figure~\ref{fig:paired-errors} shows the 40 paired comparisons. Each point is the System One model's error rate minus the control's rate in percentage points, so values below zero favor the System One model. For example, Jev's $-24.4$-point FPR difference from GPT-4.1 on AgentHarm means that it falsely flags 24.4 percentage points fewer benign requests. Lines show paired 95\% cluster-bootstrap confidence intervals, and filled markers identify differences supported by the Holm-adjusted tests described in Section~\ref{sec:design}. The counts beside model pairs give their shared input totals. FNR uses the unsafe inputs within each pair and FPR uses the benign inputs.

On WAInjectBench, none of the FNR or FPR differences is significant after adjustment at 0.05, and the small number of source clusters leaves substantial uncertainty. On AgentHarm, the evidence is clearer that Jev flags fewer benign requests than GPT-4.1, while their difference in harmful misses remains uncertain after correction.

\section{RQ2: Probability Calibration}
\label{sec:rq2}
Among inputs assigned an unsafe probability near 0.8, about 80\% should be unsafe if the predictions are calibrated. RQ2 examines this agreement across tasks and attack groups, then tests whether recalibration fitted on development data improves test probabilities.

\subsection{Probability Scores and Calibration Measures}
For System One models, we use the unsafe probabilities returned by their decision interfaces. For general judges, the score is the unsafe answer token's probability divided by the combined probabilities of the safe and unsafe tokens at the final answer position. GPT-4.1 and Qwen3-8B use separate response configurations for classification and probability analyses, so their predicted labels can differ between these analyses. Recalibration and policy evaluation use the same scoring rule across all partitions. When required labels or scores are missing, we retain the first usable retry regardless of whether its prediction is correct.

For input $i$, let $p_i$ be the unsafe probability under evaluation and $y_i$ its original label, with 1 denoting unsafe and 0 benign. Brier score is the mean of $(p_i-y_i)^2$, and negative log likelihood (NLL) is the mean binary log loss, computed after clipping probabilities to $[10^{-7},1-10^{-7}]$. Expected calibration error (ECE) compares mean probability with observed unsafe frequency in ten equal-width probability bins. Reliability diagrams plot these two values, so points above the diagonal indicate underestimated unsafe frequencies. Lower Brier, NLL, and ECE values are better. AUROC measures whether unsafe inputs receive higher scores than benign ones, whereas calibration asks whether the numerical probabilities match observed unsafe frequencies. A model can rank inputs well while reporting probabilities that are too extreme, so these measures answer different questions.

We compare the probabilities returned by each configuration, called original scores, with scores recalibrated by temperature scaling~\cite{guo2017}. After clipping, we divide the score's log odds by a temperature $T$ and convert back to a probability. We fit $T\in[0.05,20]$ by minimizing NLL on development inputs from both classes. Original and recalibrated scores are evaluated on the same inputs, with model weights unchanged.

\subsection{Calibration and High-Confidence Errors}
\begin{figure}
\centering
\includegraphics[width=\linewidth]{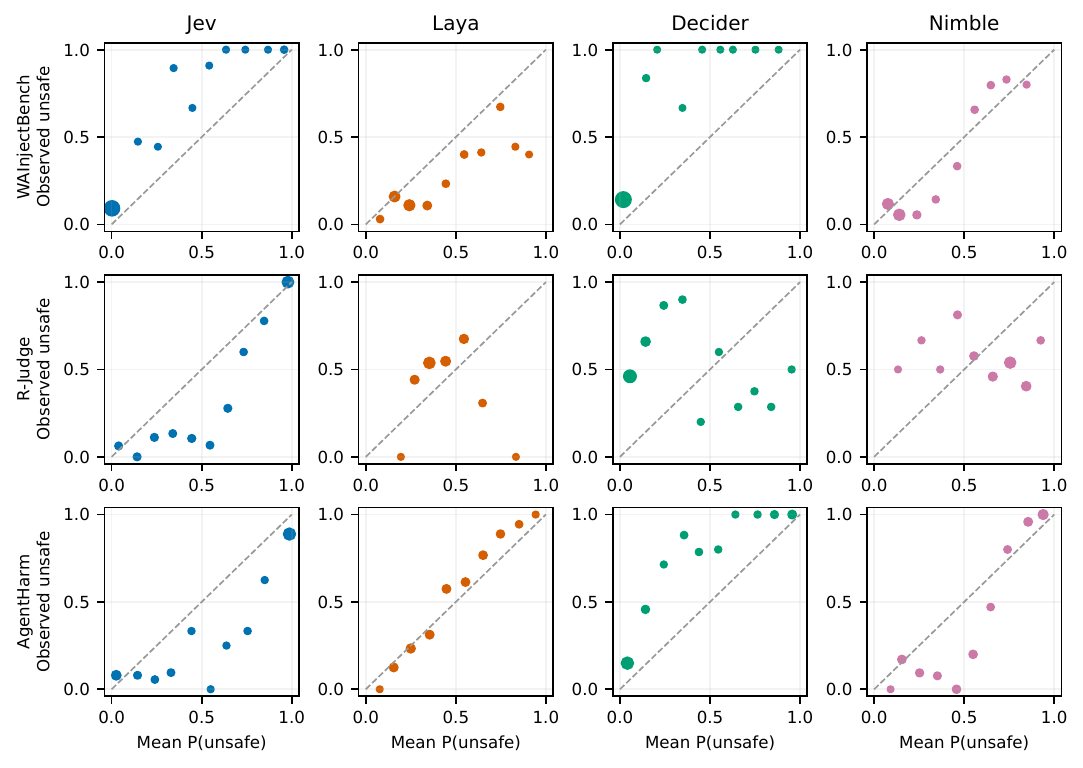}
\caption{Calibration of unsafe probabilities on test inputs.}
\label{fig:test-reliability}
\Description{Twelve panels compare predicted and empirical unsafe probabilities for four System One models on three held-out tasks, using the same test inputs for each model within a task.}
\end{figure}

Calibration summaries can hide errors in particular input groups and poor separation of safe and unsafe inputs. Table~\ref{tab:test-calibration} uses the same test inputs for every model within a task, with $N_p$ denoting the number of usable probabilities. The dagger identifies the GPT-4.1 and Qwen3-8B configurations used for probability evaluation. Figure~\ref{fig:test-reliability} compares predicted and observed unsafe frequencies across probability bins, with larger markers indicating more inputs.

\finding{Inputs assigned very low unsafe probabilities can still include many missed attacks}{finding:low-score-misses}{Jev assigns unsafe probabilities below 0.1 to 1,446 WAInjectBench inputs. Their mean predicted probability is 0.205\%, yet 133 (9.20\%) are unsafe. Of these 133 unsafe inputs, 130 lack direct attack instructions (no-EI). Removing the no-EI groups leaves an unsafe fraction of 0.228\% and a mean unsafe probability of 0.176\% among the remaining low-score inputs. Developers should check the actual unsafe fraction among inputs their policy would allow, overall and within relevant attack groups.}

\begin{table}
\caption{Probability ranking and calibration on test inputs.}
\label{tab:test-calibration}
\small
\begin{tabularx}{\linewidth}{Yrrrrrr}
\toprule
\rowcolor{tableHeader} Configuration & $N_p$ & AUROC$\uparrow$ & AP$\uparrow$ & Brier$\downarrow$ & ECE$\downarrow$ & NLL$\downarrow$ \\
\midrule
\multicolumn{7}{l}{\textbf{WAInjectBench}} \\
Nimble & 1,612 & 0.734 & 0.582 & 0.103 & 0.087 & 0.364 \\
Jev & 1,612 & 0.796 & 0.662 & 0.097 & 0.105 & 1.140 \\
ProtectAI & 1,612 & 0.797 & 0.565 & 0.139 & 0.138 & 1.234 \\
PIGuard & 1,612 & 0.657 & 0.483 & 0.132 & 0.126 & 0.933 \\
Laya & 1,612 & 0.633 & 0.345 & 0.147 & 0.108 & 0.469 \\
Nimble parent & 1,612 & 0.731 & 0.642 & 0.116 & 0.114 & 0.500 \\
Decider & 1,612 & 0.710 & 0.584 & 0.148 & 0.143 & 0.582 \\
Decider parent & 1,612 & 0.616 & 0.268 & 0.179 & 0.201 & 0.548 \\
Prompt Guard 2 & 1,612 & 0.784 & 0.427 & 0.171 & 0.171 & 1.181 \\
Qwen3-8B$^{\dagger}$ & 1,612 & 0.714 & 0.608 & 0.115 & 0.117 & 1.425 \\
GPT-4.1$^{\dagger}$ & 1,612 & 0.811 & 0.679 & 0.122 & 0.124 & 1.743 \\
GLM-5.3-Flash & 1,612 & 0.817 & 0.660 & 0.096 & 0.095 & 0.966 \\
\midrule
\multicolumn{7}{l}{\textbf{R-Judge}} \\
Jev & 236 & 0.961 & 0.975 & 0.103 & 0.137 & 0.318 \\
Decider parent & 236 & 0.570 & 0.534 & 0.258 & 0.156 & 0.714 \\
Laya & 236 & 0.533 & 0.532 & 0.272 & 0.167 & 0.741 \\
Nimble parent & 236 & 0.674 & 0.765 & 0.373 & 0.362 & 1.321 \\
Nimble & 236 & 0.437 & 0.505 & 0.320 & 0.252 & 0.875 \\
Decider & 236 & 0.610 & 0.572 & 0.423 & 0.440 & 1.262 \\
Qwen3-8B$^{\dagger}$ & 236 & 0.489 & 0.577 & 0.449 & 0.450 & 6.338 \\
GPT-4.1$^{\dagger}$ & 236 & 0.945 & 0.959 & 0.138 & 0.145 & 1.163 \\
GLM-5.3-Flash & 236 & 0.951 & 0.963 & 0.055 & 0.044 & 0.312 \\
\midrule
\multicolumn{7}{l}{\textbf{AgentHarm}} \\
Jev & 352 & 0.938 & 0.940 & 0.122 & 0.128 & 0.481 \\
Nimble & 352 & 0.935 & 0.958 & 0.123 & 0.170 & 0.390 \\
Decider & 352 & 0.918 & 0.937 & 0.153 & 0.173 & 0.497 \\
Nimble parent & 352 & 0.943 & 0.961 & 0.172 & 0.189 & 0.568 \\
Laya & 352 & 0.827 & 0.830 & 0.176 & 0.072 & 0.530 \\
WildGuard & 352 & 0.931 & 0.945 & 0.325 & 0.343 & 1.565 \\
Decider parent & 352 & 0.732 & 0.743 & 0.243 & 0.128 & 0.679 \\
Qwen3-8B$^{\dagger}$ & 352 & 0.927 & 0.934 & 0.237 & 0.242 & 2.647 \\
GPT-4.1$^{\dagger}$ & 352 & 0.931 & 0.908 & 0.238 & 0.241 & 3.244 \\
GLM-5.3-Flash & 352 & 0.818 & 0.725 & 0.160 & 0.153 & 1.159 \\
\bottomrule
\end{tabularx}
\end{table}

\textbf{Check whether unsafe inputs receive higher probabilities than benign ones.} On shared R-Judge inputs, Laya's AUROC is 0.533 versus Jev's 0.961, and its Brier score is 0.272 versus 0.103. Laya's Brier loss is 9.3\% higher than a constant-probability reference that assigns the observed test-set unsafe proportion to every input. Its scores rank the two classes little better than chance and have higher squared error than this reference.

\paragraph{Binning and constant-probability baselines.}
Motivated by Guo et al.'s calibration analysis~\cite{guo2026justask}, we check ECE with five, ten, and twenty equal-width bins (W5, W10, W20). We also use quantile bins (Q10) that aim for ten equally populated groups but may yield fewer bins because identical scores stay together. Let $\bar p$ be the mean unsafe probability and $\bar y$ the observed unsafe proportion. Table~\ref{tab:calibration-sensitivity} reports signed bias $\bar p-\bar y$ and Brier skill $1-\mathrm{Brier}/[\bar y(1-\bar y)]$. Positive bias means that mean predicted risk exceeds the observed unsafe rate. Positive Brier skill means lower squared error than assigning $\bar y$ to every input.

\begin{table}
\caption{Calibration binning and constant-probability baselines.}
\label{tab:calibration-sensitivity}
\small
\begin{tabularx}{\linewidth}{Yrrrrrr}
\toprule
\rowcolor{tableHeader} Configuration & Bias & W5$\downarrow$ & W10$\downarrow$ & W20$\downarrow$ & Q10$\downarrow$ & Skill$\uparrow$ \\
\midrule
\multicolumn{7}{l}{\textbf{WAInjectBench} ($N=1,612$, $\bar y=0.173$)} \\
Jev & -0.105 & 0.105 & 0.105 & 0.105 & 0.105 & +0.324 \\
Laya & +0.108 & 0.108 & 0.108 & 0.116 & 0.115 & -0.024 \\
Decider & -0.143 & 0.143 & 0.143 & 0.143 & 0.143 & -0.031 \\
Nimble & +0.036 & 0.052 & 0.087 & 0.088 & 0.090 & +0.282 \\
\midrule
\multicolumn{7}{l}{\textbf{R-Judge} ($N=236$, $\bar y=0.530$)} \\
Jev & +0.116 & 0.129 & 0.137 & 0.145 & 0.132 & +0.588 \\
Laya & -0.119 & 0.167 & 0.167 & 0.167 & 0.162 & -0.093 \\
Decider & -0.334 & 0.438 & 0.440 & 0.442 & 0.428 & -0.697 \\
Nimble & +0.168 & 0.252 & 0.252 & 0.265 & 0.253 & -0.286 \\
\midrule
\multicolumn{7}{l}{\textbf{AgentHarm} ($N=352$, $\bar y=0.500$)} \\
Jev & +0.105 & 0.119 & 0.128 & 0.130 & 0.131 & +0.510 \\
Laya & -0.045 & 0.072 & 0.072 & 0.085 & 0.089 & +0.295 \\
Decider & -0.173 & 0.173 & 0.173 & 0.173 & 0.173 & +0.387 \\
Nimble & +0.101 & 0.162 & 0.170 & 0.172 & 0.167 & +0.507 \\
\bottomrule
\end{tabularx}
\end{table}

Across bin choices, Jev's WAInjectBench ECE changes little, and Laya's R-Judge ECE is unchanged across the three equal-width choices. Laya's near-chance R-Judge AUROC indicates weak ranking of unsafe above benign inputs. Brier skill is negative for Laya and Decider on WAInjectBench and for Laya, Decider, and Nimble on R-Judge. Their probabilities have higher squared error than assigning every input the test set's unsafe proportion, even though that constant reference provides no information about which individual inputs are unsafe.

\begin{figure}
\centering
\includegraphics[width=\linewidth]{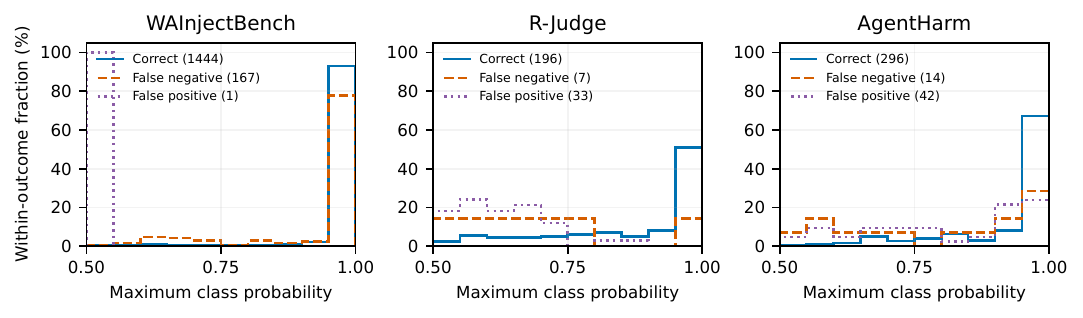}
\caption{Maximum class probability by decision outcome for Jev.}
\label{fig:confident-errors}
\Description{Three panels show normalized confidence histograms for correct decisions, false negatives, and false positives on the same test inputs.}
\end{figure}

For confidence comparisons, we use the larger class probability $\max(p_i,1-p_i)$. At values $\geq0.95$, WAInjectBench false-negative counts are 130 for Jev, 162 for Decider, one for Nimble, and zero for Laya. Jev instead makes four false negatives and ten false positives at this confidence on AgentHarm. Laya reaches this confidence on only three WAInjectBench inputs, all correct.
Figure~\ref{fig:confident-errors} shows maximum class probability separately for correct decisions, false negatives, and false positives, with each histogram normalized within its outcome group.

\paragraph{Reasoning-judge probabilities.}
The reasoning judges also show that ranking unsafe inputs above safe ones does not ensure accurate probabilities. Table~\ref{tab:test-calibration} includes GLM-5.3-Flash, while Table~\ref{tab:added-calibration} reports the remaining reasoning judges on the same test inputs. Kimi's AgentHarm AUROC of 0.922 and AP of 0.909 indicate strong ranking, but its Brier and ECE are both 0.273, and it falsely flags 96 benign requests. DeepSeek's R-Judge Brier loss of 0.034 accompanies the highest Macro-F1 point estimate in that task's reasoning-judge comparison. For the intended task, compare missed unsafe inputs and benign false alarms together with calibration metrics.

\begin{table}
\caption{Probability ranking and calibration of reasoning judges.}
\label{tab:added-calibration}
\small
\begin{tabularx}{\linewidth}{Yrrrrr}
\toprule
\rowcolor{tableHeader} Configuration & AUROC$\uparrow$ & AP$\uparrow$ & Brier$\downarrow$ & ECE$\downarrow$ & NLL$\downarrow$ \\
\midrule
\multicolumn{6}{l}{\textbf{WAInjectBench}} \\
DeepSeek-V4.1-Flash & 0.774 & 0.626 & 0.099 & 0.099 & 1.598 \\
GLM-5.3 & 0.685 & 0.531 & 0.116 & 0.117 & 1.187 \\
Kimi-K2.6 & 0.786 & 0.642 & 0.122 & 0.122 & 1.580 \\
Qwen3.8-27B & 0.645 & 0.512 & 0.112 & 0.112 & 1.485 \\
\midrule
\multicolumn{6}{l}{\textbf{R-Judge}} \\
DeepSeek-V4.1-Flash & 0.968 & 0.959 & 0.034 & 0.034 & 0.546 \\
GLM-5.3 & 0.958 & 0.974 & 0.067 & 0.073 & 0.444 \\
Kimi-K2.6 & 0.937 & 0.920 & 0.093 & 0.093 & 0.856 \\
Qwen3.8-27B & 0.954 & 0.971 & 0.055 & 0.055 & 0.759 \\
\midrule
\multicolumn{6}{l}{\textbf{AgentHarm}} \\
DeepSeek-V4.1-Flash & 0.877 & 0.844 & 0.213 & 0.213 & 3.433 \\
GLM-5.3 & 0.813 & 0.710 & 0.156 & 0.131 & 1.015 \\
Kimi-K2.6 & 0.922 & 0.909 & 0.273 & 0.273 & 2.473 \\
Qwen3.8-27B & 0.713 & 0.619 & 0.239 & 0.239 & 3.031 \\
\bottomrule
\end{tabularx}
\end{table}

\subsection{Comparing Adapted Models with Their Base Models}
\begin{figure}
\centering
\includegraphics[width=\linewidth]{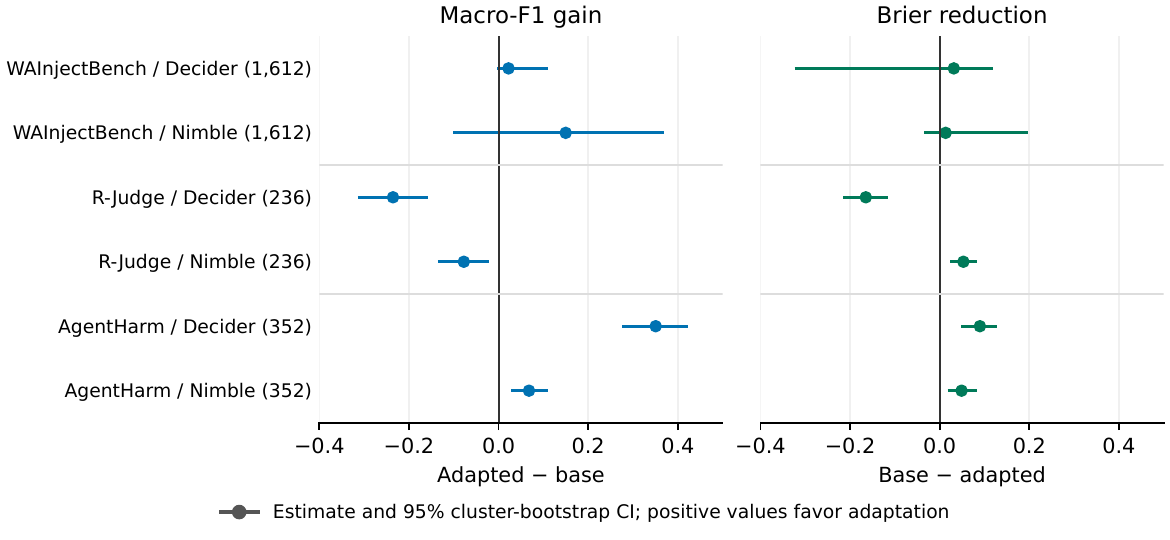}
\caption{Classification and probability differences from base configurations.}
\label{fig:parent-comparison}
\Description{Two panels show Macro-F1 gains and Brier reductions for Decider and Nimble relative to their base-model controls on three tasks. Both axes use positive values for improvements, with paired 95 percent cluster-bootstrap confidence intervals.}
\end{figure}

We compare classification and probability metrics for Decider and Nimble against their respective base configurations. Figure~\ref{fig:parent-comparison} shows all twelve paired comparisons on the complete test inputs for each pair, with 95\% cluster-bootstrap confidence intervals. Both panels put improvements on the right: Macro-F1 gain is adapted minus base, while Brier reduction is base minus adapted. Both adapted configurations have higher Macro-F1 and lower Brier loss than their base controls on AgentHarm. On R-Judge, Nimble has lower Macro-F1 by 0.078 and lower Brier loss by 0.053, while Decider is worse on both metrics. The adapted and base configurations differ in both weights and scoring temperature (Section~\ref{sec:design}).

\subsection{Effects of Recalibration}
For each model, we compare original and recalibrated scores on the same complete test sets used in the cross-model tables. Temperature scaling improves test NLL in six of the twelve System One model and task combinations, including all four R-Judge comparisons, but the gains do not extend to every metric or task. On Jev's 1,612 valid WAInjectBench outputs, NLL decreases from 1.140 to 0.374 while Brier increases from 0.097 to 0.107. On AgentHarm, Jev and Nimble classify all 64 development inputs correctly, leading the fit to select a temperature near the lower bound of 0.05 for both. This development fit makes probabilities more extreme, increasing the penalty for high-confidence test errors and raising Jev's test NLL from 0.481 to 2.163.

Temperature scaling reduces test NLL in all six GPT-4.1 and Qwen3-8B task comparisons using the evaluated scoring configurations. On all R-Judge inputs with usable scores, Qwen's NLL falls from 6.338 to 0.730 because scaling softens its often near-zero or near-one probabilities, lowering the loss from high-confidence errors.
Table~\ref{tab:extended-temperatures} lists the original and recalibrated results, in that order, within each metric cell.

\begin{table}
\caption{Test calibration before and after temperature scaling.}
\label{tab:extended-temperatures}
\small
\begin{tabularx}{\linewidth}{Yrrrr}
\toprule
\rowcolor{tableHeader} Configuration & $T$ & Brier$\downarrow$ & ECE$\downarrow$ & NLL$\downarrow$ \\
\midrule
\multicolumn{5}{l}{\textbf{WAInjectBench}} \\
Jev & 10.93 & 0.097 / 0.107 & 0.105 / 0.133 & 1.140 / 0.374 \\
Laya & 15.19 & 0.147 / 0.237 & 0.108 / 0.310 & 0.469 / 0.666 \\
Decider & 20.00 & 0.148 / 0.216 & 0.143 / 0.285 & 0.582 / 0.625 \\
Nimble & 1.18 & 0.103 / 0.107 & 0.087 / 0.104 & 0.364 / 0.376 \\
\midrule
\multicolumn{5}{l}{\textbf{R-Judge}} \\
Jev & 0.90 & 0.103 / 0.102 & 0.137 / 0.131 & 0.318 / 0.316 \\
Laya & 20.00 & 0.272 / 0.250 & 0.167 / 0.034 & 0.741 / 0.693 \\
Decider & 20.00 & 0.423 / 0.251 & 0.440 / 0.087 & 1.262 / 0.695 \\
Nimble & 20.00 & 0.320 / 0.251 & 0.252 / 0.034 & 0.875 / 0.694 \\
\midrule
\multicolumn{5}{l}{\textbf{AgentHarm}} \\
Jev & 0.05 & 0.122 / 0.156 & 0.128 / 0.158 & 0.481 / 2.163 \\
Laya & 1.09 & 0.176 / 0.178 & 0.072 / 0.077 & 0.530 / 0.535 \\
Decider & 1.67 & 0.153 / 0.145 & 0.173 / 0.140 & 0.497 / 0.453 \\
Nimble & 0.05 & 0.123 / 0.160 & 0.170 / 0.169 & 0.390 / 1.454 \\
\bottomrule
\end{tabularx}
\end{table}

Temperature scaling makes probabilities more or less extreme, but inputs with the same original score still receive the same transformed score. A threshold must therefore handle these tied inputs together, even when some decisions are correct and others are wrong. RQ3 examines how this limits automatic decisions.

\section{RQ3: Automation within Error Limits}
\label{sec:rq3}
RQ3 examines how many inputs a policy can process without review while limiting how often it allows unsafe inputs or blocks benign ones. These limits, called error budgets, are checked using benchmark labels on confirmation and test inputs after thresholds are fixed. We also compare model errors to assess potential reviewers.

\subsection{Decision Policy and Error Measures}
The policy compares unsafe probability $p_i$ with an allow threshold $\tau_L$ and a block threshold $\tau_H$:
\begin{equation}
 \text{Decision}=
 \begin{cases}
 \Allow, & p_i<\tau_L,\\
 \Block, & p_i>\tau_H,\\
 \Escalate, & p_i\in[\tau_L,\tau_H].
 \end{cases}
 \label{eq:policy}
\end{equation}
Inputs exactly at either threshold, or without usable scores, also escalate for review. Automatic coverage is the fraction allowed or blocked, while allow coverage counts only allowances. Both use all partition inputs as the denominator. We report them separately because high coverage can result from blocking many inputs while allowing few to proceed. The unsafe-miss rate $\Miss$ is the fraction of unsafe inputs allowed, and the benign-blocking rate $\FalseBlock$ is the fraction of benign inputs blocked. A low miss rate can therefore come from allowing almost nothing.
Allowed-set risk instead measures the unsafe fraction among allowed inputs, with zero-denominator ratios undefined.

\subsection{Policy Selection and Independent Confirmation}
Symmetric policies link the thresholds through $\tau_L=1-\tau_H$. We evaluate 101 pairs with $\tau_H$ increasing from 0.5 to 1 in steps of 0.005. For each allowed miss rate $\epsilon\in\{0.01,0.05,0.10\}$, we choose the pair with the highest automatic coverage on selection inputs, subject to $\Miss\leq\epsilon$ and $\FalseBlock\leq0.05$. If coverage is tied, we prefer more automatic allowances, then the larger $\tau_H$. The search includes $(\tau_L,\tau_H)=(0,1)$, which sends every input for review and automates no decisions.

Selected thresholds remain fixed for confirmation and test, including policies that fail confirmation. A policy passes only when both error rates meet their limits. These checks use WAInjectBench and R-Judge, which provide the required partitions (Table~\ref{tab:offline-splits}).

\paragraph{Small samples and strict error limits.}
A 1\% miss limit requires zero misses among R-Judge's 74 unsafe selection inputs and 49 unsafe confirmation inputs, because a single miss exceeds the limit in either set. WAInjectBench has 167 and 111, respectively, permitting one miss in each. With so few permitted errors, individual inputs can determine whether a policy passes, and larger samples are needed to estimate rare-error rates precisely.
The benign-blocking limit is similarly sensitive to individual errors. R-Judge confirmation permits at most two blocks among 44 benign inputs under its 5\% limit, while test permits five among 111. Cluster-bootstrap intervals describe uncertainty from the sampled interaction records and attack sources.

\subsection{Coverage and Errors of Symmetric Policies}
Table~\ref{tab:test-policy} reports test results at the 1\% selection miss limit. A ``Pass'' under ``Confirmation check'' means that the 1\% unsafe-miss and 5\% benign-blocking limits were both met on separate confirmation inputs. The double dagger identifies the GPT-4.1 and Qwen3-8B probability-scoring configurations used for threshold selection.
\begin{table}
\caption{Symmetric thresholds at the 1\% selection miss limit.}
\label{tab:test-policy}
\small
\begin{tabularx}{\linewidth}{Yrrrrc}
\toprule
\rowcolor{tableHeader} Configuration & Coverage & Allowed & Misses & \shortstack{False\\blocks} & \shortstack{Confirmation\\check} \\
\midrule
\multicolumn{6}{l}{\textbf{WAInjectBench}} \\
Jev & 0.00\% & 0.00\% & 0.00\% & 0.00\% & Pass \\
Laya & 5.02\% & 4.71\% & 1.08\% & 0.23\% & Pass \\
Decider & 0.00\% & 0.00\% & 0.00\% & 0.00\% & Pass \\
Nimble & 0.74\% & 0.74\% & 0.36\% & 0.00\% & Pass \\
Qwen3-8B$^{\ddagger}$ & 0.00\% & 0.00\% & 0.00\% & 0.00\% & Pass \\
GPT-4.1$^{\ddagger}$ & 0.00\% & 0.00\% & 0.00\% & 0.00\% & Pass \\
\midrule
\multicolumn{6}{l}{\textbf{R-Judge}} \\
Jev & 7.63\% & 0.42\% & 0.00\% & 0.00\% & Pass \\
Laya & 1.69\% & 1.27\% & 0.00\% & 0.90\% & Fail \\
Decider & 1.69\% & 1.27\% & 0.80\% & 0.00\% & Pass \\
Nimble & 6.78\% & 0.42\% & 0.00\% & 6.31\% & Pass \\
Qwen3-8B$^{\ddagger}$ & 0.00\% & 0.00\% & 0.00\% & 0.00\% & Pass \\
GPT-4.1$^{\ddagger}$ & 0.00\% & 0.00\% & 0.00\% & 0.00\% & Pass \\
\bottomrule
\end{tabularx}
\end{table}

With a 1\% miss limit during selection, symmetric policies send every WAInjectBench input for review with Jev, Decider, GPT-4.1, and Qwen3-8B (Table~\ref{tab:test-policy}). Laya automates 5.02\% but allows three of 279 unsafe inputs, exceeding the miss limit. On R-Judge, Nimble passes confirmation but blocks seven of 111 benign test inputs, exceeding the 5\% blocking limit. Jev's 7.63\% R-Judge coverage consists of 17 blocks and one allowance, so only 0.42\% of inputs proceed without review. GPT-4.1 and Qwen3-8B escalate every R-Judge input with either original or recalibrated scores. These policies show that passing confirmation does not ensure the error limits hold on test inputs, and lower NLL after recalibration need not produce automatic decisions.

\begin{figure}
\centering
\includegraphics[width=\linewidth]{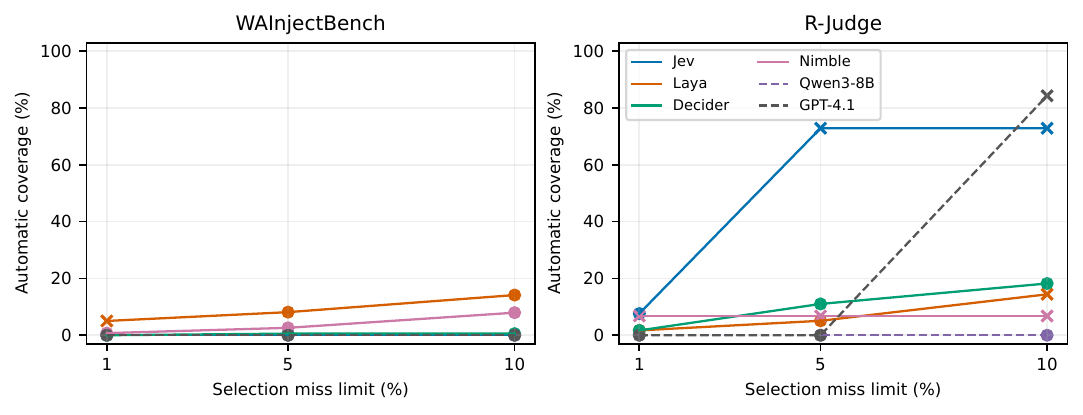}
\caption{Automatic coverage at different selection miss limits.}
\label{fig:test-automation}
\Description{Two panels show automatic coverage at three source miss targets for four System One models, GPT-4.1, and Qwen3-8B. Dashed curves denote the judges.}
\end{figure}

Raising the miss limit used for threshold selection can increase coverage but also change benign blocking (Figure~\ref{fig:test-automation}). Circles mark policies meeting both test limits, and crosses mark failures. Jev's R-Judge coverage rises from 7.63\% to 72.88\% at a 5\% miss budget, but benign blocking reaches 5.41\%, above its unchanged limit. This policy allows 54 inputs, including three unsafe ones. Those misses are 2.4\% of all unsafe inputs but 5.56\% of allowances, illustrating why low overall miss rates need not imply low risk among allowed inputs.

\subsection{Independent Allowance and Blocking Thresholds}
\label{sec:threshold-sensitivity}
We examine independent allow and block thresholds by removing the symmetry constraint: choosing a low allow threshold no longer forces a high block threshold. Each threshold has 101 evenly spaced candidates within $\tau_L\in[0,0.5]$ and $\tau_H\in[0.5,1]$, giving $101\times101=10{,}201$ possible pairs. We separately maximize allowances within the miss limit and blocks within the benign-blocking limit. Because the score ranges do not overlap, this also maximizes total coverage. We use the same tie-breaking rules, then prefer the smaller $\tau_L$ if a tie remains. We select thresholds on selection inputs and apply them unchanged to confirmation and test, comparing both policy types with original and recalibrated scores at all three miss limits.

\begin{figure}
\centering
\includegraphics[width=\linewidth]{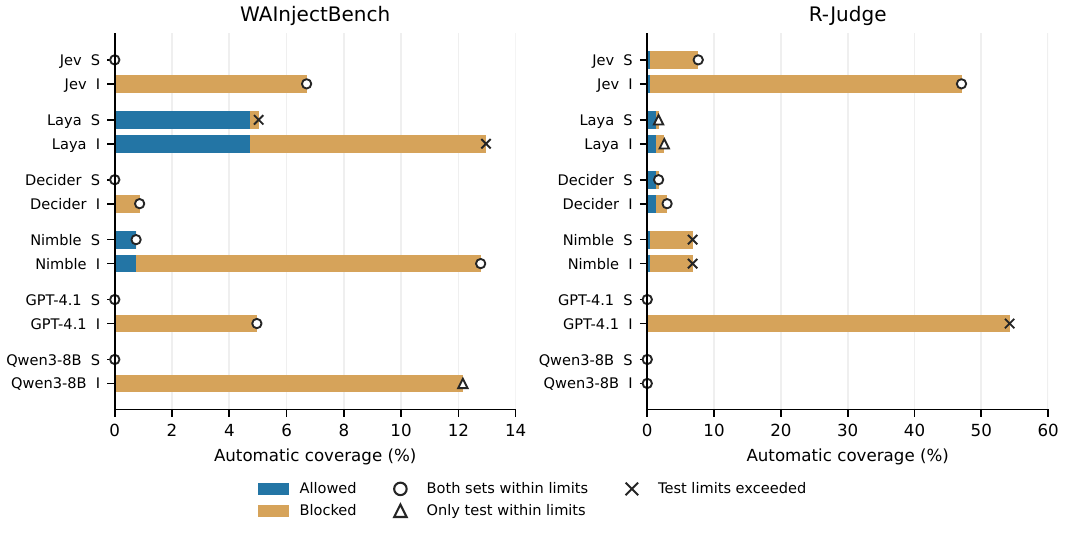}
\caption{Automation with symmetric and independent thresholds.}
\label{fig:threshold-sensitivity}
\Description{Paired stacked bars show percentages of test inputs allowed in blue and blocked in yellow under symmetric and independent thresholds. Circles identify policies meeting both the one-percent unsafe-miss and five-percent benign-blocking limits on confirmation and test. Triangles identify policies meeting both limits on test only. Crosses identify policies exceeding at least one limit on test.}
\end{figure}

Figure~\ref{fig:threshold-sensitivity} compares symmetric (S) and independent (I) thresholds using original scores. Blue and yellow bars show the percentages of test inputs allowed and blocked. Circles mean both confirmation and test meet the 1\% unsafe-miss and 5\% benign-blocking limits. Triangles mean only test meets both limits. Crosses mean test exceeds at least one limit.

\finding{Selecting allow and block thresholds independently can increase the number of blocked inputs while leaving the number of allowed inputs unchanged}{finding:frozen-automation}{On R-Judge, independent thresholds raise Jev's test coverage from 7.63\% to 47.03\% while meeting confirmation and test limits. Both policies allow just one input, while independent thresholds block 110. On WAInjectBench, coverage rises from zero to 6.70\%, entirely through 108 blocks. Developers should report allowed and blocked fractions separately and compare observed error rates with both limits on independent labeled data.}

Nimble similarly rises from 0.74\% to 12.78\% WAInjectBench coverage while allowing the same twelve inputs. Independent thresholds separate the two actions, but higher selection coverage does not ensure that error limits hold on other inputs. GPT-4.1 blocks 128 R-Judge inputs, including 17 benign ones, and fails both confirmation and test checks. Among the twelve model/task pairs with original scores at the 1\% miss limit, seven independent policies pass both checks, compared with nine symmetric policies. Several passing symmetric policies send every input for review.

\begin{table}
\caption{Independent thresholds at the 1\% selection miss limit.}
\label{tab:independent-policy}
\small
\begin{tabularx}{\linewidth}{Yrrrrrrcc}
\toprule
\rowcolor{tableHeader} Configuration & $\tau_L$ & $\tau_H$ & Allowed & Blocked & Misses & \shortstack{False\\blocks} & \shortstack{Confirmation\\check} & \shortstack{Test\\check} \\
\midrule
\multicolumn{9}{l}{\textbf{WAInjectBench}} \\
Jev & 0.000 & 0.525 & 0 & 108 & 0 & 0 & Pass & Pass \\
Laya & 0.105 & 0.570 & 76 & 133 & 3 & 65 & Pass & Fail \\
Decider & 0.000 & 0.515 & 0 & 14 & 0 & 0 & Pass & Pass \\
Nimble & 0.050 & 0.500 & 12 & 194 & 1 & 47 & Pass & Pass \\
GPT-4.1 & 0.000 & 0.655 & 0 & 80 & 0 & 2 & Pass & Pass \\
Qwen3-8B & 0.000 & 0.560 & 0 & 196 & 0 & 54 & Fail & Pass \\
\midrule
\multicolumn{9}{l}{\textbf{R-Judge}} \\
Jev & 0.005 & 0.745 & 1 & 110 & 0 & 2 & Pass & Pass \\
Laya & 0.220 & 0.685 & 3 & 3 & 0 & 3 & Fail & Pass \\
Decider & 0.015 & 0.925 & 3 & 4 & 1 & 2 & Pass & Pass \\
Nimble & 0.150 & 0.885 & 1 & 15 & 0 & 7 & Pass & Fail \\
GPT-4.1 & 0.000 & 0.985 & 0 & 128 & 0 & 17 & Fail & Fail \\
Qwen3-8B & 0.000 & 1.000 & 0 & 0 & 0 & 0 & Pass & Pass \\
\bottomrule
\end{tabularx}
\par\noindent Allowed, blocked, misses, and false blocks are test-input counts. Pass means both error limits are met.
\end{table}

Table~\ref{tab:independent-policy} reports test counts and separate confirmation and test checks against both error limits. Both Laya policies allow the same 76 inputs, including three unsafe ones. Independent thresholds add 128 blocks, including 62 benign inputs, so the higher coverage leaves the missed attacks unchanged while rejecting more benign inputs.

\begin{table}
\caption{Independent thresholds after probability recalibration.}
\label{tab:independent-scaled}
\small
\begin{tabularx}{\linewidth}{Yrrrrrrrcc}
\toprule
\rowcolor{tableHeader} Configuration & $\tau_L$ & $\tau_H$ & \multicolumn{2}{c}{Coverage (\%)} & \multicolumn{3}{c}{Independent policy (\%)} & \multicolumn{2}{c}{Budget check} \\
\cmidrule(lr){4-5}\cmidrule(lr){6-8}\cmidrule(lr){9-10}
\rowcolor{tableHeader} & & & S & I & Allowed & Misses & \shortstack{False\\blocks} & Confirmation & Test \\
\midrule
\multicolumn{10}{l}{\textbf{WAInjectBench}} \\
Jev & 0.000 & 0.500 & 0.00 & 6.95 & 0.00 & 0.00 & 0.08 & Pass & Pass \\
Laya & 0.465 & 0.505 & 5.15 & 12.90 & 4.84 & 1.08 & 4.73 & Pass & Fail \\
Decider & 0.000 & 0.500 & 0.00 & 0.87 & 0.00 & 0.00 & 0.00 & Pass & Pass \\
Nimble & 0.075 & 0.500 & 0.37 & 12.41 & 0.37 & 0.00 & 3.53 & Pass & Pass \\
GPT-4.1 & 0.000 & 0.505 & 0.00 & 4.96 & 0.00 & 0.00 & 0.15 & Pass & Pass \\
Qwen3-8B & 0.000 & 0.505 & 0.00 & 12.16 & 0.00 & 0.00 & 4.05 & Fail & Pass \\
\midrule
\multicolumn{10}{l}{\textbf{R-Judge}} \\
Jev & 0.005 & 0.770 & 7.63 & 47.03 & 0.42 & 0.00 & 1.80 & Pass & Pass \\
Laya & 0.485 & 0.510 & 1.69 & 2.54 & 1.27 & 0.00 & 2.70 & Fail & Pass \\
Decider & 0.000 & 0.530 & 0.00 & 1.69 & 0.00 & 0.00 & 1.80 & Pass & Pass \\
Nimble & 0.480 & 0.530 & 2.12 & 2.97 & 0.85 & 0.80 & 1.80 & Pass & Pass \\
GPT-4.1 & 0.000 & 0.625 & 0.00 & 54.24 & 0.00 & 0.00 & 15.32 & Fail & Fail \\
Qwen3-8B & 0.330 & 1.000 & 0.00 & 1.69 & 1.69 & 0.00 & 0.00 & Fail & Pass \\
\bottomrule
\end{tabularx}
\par\noindent S/I: symmetric/independent thresholds. All rates are measured on test inputs. Pass means both error limits are met.
\end{table}

At the 1\% selection miss limit, recalibrated scores also gain coverage without consistently meeting both error limits (Table~\ref{tab:independent-scaled}). On R-Judge, Jev again reaches 47.03\% coverage with one allowance, and Decider blocks four inputs that its symmetric policy sent for review. Qwen3-8B gains four allowances but fails confirmation. Across miss limits of 1\%, 5\%, and 10\%, independent policies pass both checks in 7, 9, and 7 of the twelve recalibrated model/task pairs, compared with 10, 9, and 9 symmetric policies.

For Jev's original scores on R-Judge, independent thresholds increase test coverage by 39.41 percentage points (95\% cluster-bootstrap CI $[33.05,45.34]$), entirely from additional blocks. The interval describes variation across test samples with the selected thresholds fixed.

\subsection{Confidence Ranking and Identical Scores}
\begin{figure}
\centering
\includegraphics[width=\linewidth]{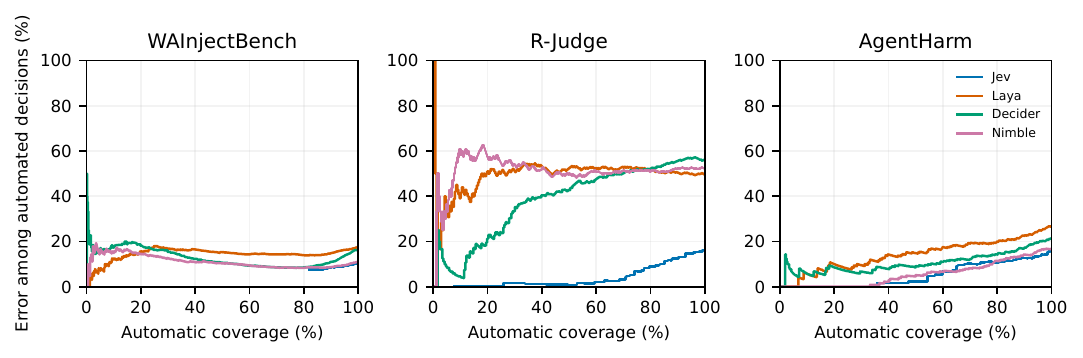}
\caption{Error rate versus automatic coverage.}
\label{fig:selective-risk}
\Description{Three panels show error among accepted decisions versus coverage for four System One models. Equal-confidence inputs enter together.}
\end{figure}

Figure~\ref{fig:selective-risk} ranks inputs by maximum class probability and counts predictions as automatic decisions from highest to lowest confidence. A safe prediction allows the input and an unsafe prediction blocks it. Inputs with equal confidence enter together. The plotted error rate is the fraction of these decisions that are wrong.

\textbf{Thresholds cannot separate correct and incorrect decisions with the same score.} Jev assigns $p_i=0$ to 1,320 shared WAInjectBench inputs, including 102 misses, and $p_i=1$ to three correct predictions. Automatically allowing the zero-score inputs and blocking the one-score inputs covers 82.07\% at 7.71\% error. Any positive allow threshold admits all zero-score inputs, including the missed attacks. An independent block threshold can reject high-score inputs but cannot distinguish attacks from benign inputs within that zero-score group. On AgentHarm, Jev's predictions with maximum class probability one cover 126 of 352 inputs with two errors (1.59\%), showing that the same confidence accompanies different observed error rates across tasks.

\subsection{Language-Model Judges as Potential Reviewers}
\label{sec:complementarity}
An application can send uncertain inputs to a language-model judge for review. To assess models for this second-judge role, we compare their predictions with those of each System One model on the same inputs~\cite{rao2026rubricjudges}. We count an error as corrected when the judge predicts the benchmark label for an input the System One model misclassifies. A judge can correct the first model's missed attack only when the review rule forwards that input, so selecting the reviewer also requires checking which cases the rule leaves unreviewed.

\begin{figure}
\centering
\includegraphics[width=\linewidth]{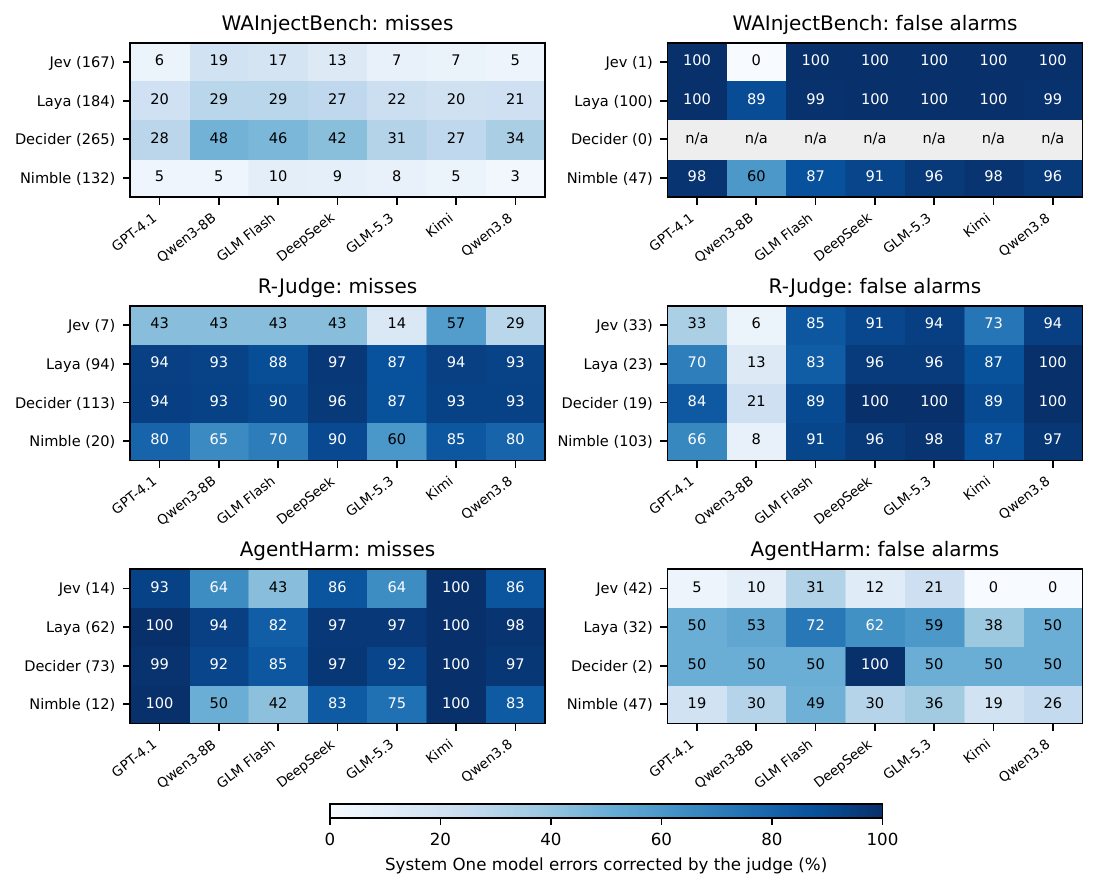}
\caption{System One model errors corrected by language-model judges.}
\label{fig:error-complementarity}
\Description{Six heatmaps show the fraction of false negatives or false positives of four System One models corrected by each of seven general judges. Rows report the base error count.}
\end{figure}

We compare four System One models with seven general judges on 1,612 shared WAInjectBench inputs, 236 R-Judge records, and 352 AgentHarm requests. Including task-compatible specialists gives 100 model pairs evaluated on the same complete test inputs within each task. For each class, a judge can correct or repeat the System One model's errors, or make errors on inputs that model classified correctly. Figure~\ref{fig:error-complementarity} reports the fraction of the System One model's errors corrected by each judge. Row counts give the original errors, and grey cells indicate no errors to correct. Table~\ref{tab:error-transitions} also counts errors on inputs Jev classified correctly. GPT-4.1 identifies 10 of the 167 WAInjectBench attacks missed by Jev but misses 33 that Jev correctly flags. Its own miss count is therefore 190, showing that correcting some Jev errors need not reduce the total misses.

\begin{table}
\caption{Errors corrected and added by judges compared with Jev.}
\label{tab:error-transitions}
\small
\begin{tabularx}{\linewidth}{Yrrrr}
\toprule
\rowcolor{tableHeader} Judge & \multicolumn{2}{c}{False negatives} & \multicolumn{2}{c}{False positives} \\
\cmidrule(lr){2-3}\cmidrule(lr){4-5}
\rowcolor{tableHeader} & Corrected & Added & Corrected & Added \\
\midrule
\multicolumn{5}{l}{\textbf{WAInjectBench, $N=1,612$}} \\
GPT-4.1 & 10 of 167 & 33 & 1 of 1 & 2 \\
Qwen3-8B & 32 of 167 & 3 & 0 of 1 & 51 \\
GLM-5.3-Flash & 29 of 167 & 5 & 1 of 1 & 14 \\
DeepSeek-V4.1-Flash & 22 of 167 & 8 & 1 of 1 & 7 \\
GLM-5.3 & 12 of 167 & 29 & 1 of 1 & 6 \\
Kimi-K2.6 & 11 of 167 & 38 & 1 of 1 & 3 \\
Qwen3.8-27B & 8 of 167 & 16 & 1 of 1 & 6 \\
\midrule
\multicolumn{5}{l}{\textbf{R-Judge, $N=236$}} \\
GPT-4.1 & 3 of 7 & 4 & 11 of 33 & 14 \\
Qwen3-8B & 3 of 7 & 4 & 2 of 33 & 69 \\
GLM-5.3-Flash & 3 of 7 & 8 & 28 of 33 & 4 \\
DeepSeek-V4.1-Flash & 3 of 7 & 0 & 30 of 33 & 1 \\
GLM-5.3 & 1 of 7 & 10 & 31 of 33 & 1 \\
Kimi-K2.6 & 4 of 7 & 5 & 24 of 33 & 5 \\
Qwen3.8-27B & 2 of 7 & 5 & 31 of 33 & 1 \\
\midrule
\multicolumn{5}{l}{\textbf{AgentHarm, $N=352$}} \\
GPT-4.1 & 13 of 14 & 0 & 2 of 42 & 45 \\
Qwen3-8B & 9 of 14 & 1 & 4 of 42 & 40 \\
GLM-5.3-Flash & 6 of 14 & 8 & 13 of 42 & 19 \\
DeepSeek-V4.1-Flash & 12 of 14 & 2 & 5 of 42 & 34 \\
GLM-5.3 & 9 of 14 & 5 & 9 of 42 & 27 \\
Kimi-K2.6 & 14 of 14 & 0 & 0 of 42 & 54 \\
Qwen3.8-27B & 12 of 14 & 0 & 0 of 42 & 40 \\
\bottomrule
\end{tabularx}
\end{table}

On WAInjectBench, Jev misses 167 unsafe inputs, and the seven judges correct between 8 and 32. Among misses with $\max(p_i,1-p_i)\geq0.95$, each judge detects at most five of Jev's 130 misses and 30 of Decider's 162. Each judge repeats over 80\% of these errors for both models, and a rule accepting predictions at that confidence would send none of these misses for review. All seven judges also repeat Nimble's single high-confidence miss, while Laya has no such misses.

On AgentHarm, some judges detect previously missed harmful requests but falsely flag more benign ones. Kimi corrects all 14 Jev misses without introducing another miss, but repeats all 42 Jev false positives and adds 54. GPT-4.1 corrects 13 misses and introduces none, while correcting two false positives and adding 45. On shared R-Judge inputs, DeepSeek corrects 30 of Jev's 33 false positives and three of its seven misses, introducing one false positive and no new miss.

Across model and task pairs, every judge corrects over 80\% of Laya's and Decider's AgentHarm misses, while no judge corrects even 10\% of Nimble's WAInjectBench misses. Choosing a review judge therefore requires checking the errors it corrects and introduces, along with which inputs reach it. Reviewing only uncertain inputs leaves confident mistakes unreviewed, even when another judge could detect them. Developers should apply the review rule to labeled validation inputs and count how many missed unsafe inputs it forwards, then compare both judges' misses and benign false alarms on those inputs.

\section{Discussion}
\label{sec:discussion}

\subsection{Component Integration and Error Handling}
Applications need explicit handling of input limits, invalid answers, missing scores, and network failures, sending inputs without usable scores for review. Deployment measurements also depend on the configuration: service routing, queueing, caching, and reasoning settings affect latency and API charges, while local backends and precision can affect scores.

\subsection{Implications for Developing Agentic Software}

\textbf{Define the security task before selecting a model.} Developers should specify the behavior to prevent, the legitimate work to preserve, and where the check affects the workflow. Checks on retrieved content, interactions, and user requests need different inputs and labels. For example, a request to summarize a dangerous command differs from a request to execute it, so the question, answer definitions, and available context should make the intended judgment explicit.

\textbf{Choose models for the application's requirements.} Developers should compare candidates on representative labeled inputs at tolerable miss and false-alarm rates. The comparison should include task-compatible specialists alongside reusable typed components and general judges, taking context capacity and deployment resources into account. Because rankings and adaptation gains vary by task, candidates need to be evaluated with the question and input context the application will actually use.

\textbf{Validate labels, probabilities, and policy actions separately.} Developers should check classification errors across relevant input and attack groups, then assess whether probabilities reflect observed outcomes, particularly among inputs that would be allowed. Recalibration should be fitted on development inputs and cutoffs selected on selection inputs, with both frozen before independent checks. Policy validation should count unsafe inputs allowed, benign inputs blocked, and inputs sent for review, reporting allowed and blocked fractions separately because additional blocks do not enable more work to proceed. This staged validation supports component testing and monitoring in production ML engineering~\cite{amershi2019,breck2017}.

\textbf{Choose which inputs to review and which model reviews them.} The second judge receives the inputs selected for review, which can differ substantially from the full workload. Developers should use labeled inputs to measure which missed unsafe inputs the rule forwards and which errors the judge corrects or introduces. If validation reveals confident mistakes that an uncertainty-based rule leaves unreviewed, developers should reconsider the question, context, model, or review rule, then test the revised combination.

\textbf{Maintain validation as the software changes.} Developers should record the model version, question, input construction, and cutoffs as one tested configuration and revalidate it whenever any part changes. Monitoring should include automatically handled inputs as well as reviewed ones, because errors in the former may never reach the second judge. Applications should define an explicit fallback for configurations that fail these checks.

\textbf{Enforce permissions independently of model judgments.} Access controls should restrict the resources an agent can use, while deterministic runtime checks enforce permitted operations~\cite{camel,agentspec2026}. A safe prediction should not override these permissions, and the combined system should be evaluated by measuring successful attacks and completion of legitimate tasks.

\section{Threats to Validity}
\label{sec:validity}
\textbf{Internal validity.} Comparisons depend on prompts, answer-label order, context limits, reasoning settings, and services. Response sampling and output formats can affect scores. Recalibration uses only development data, and threshold selection uses only selection data, with no retuning on confirmation or test. Rare errors and few source clusters limit interval precision.
\textbf{Construct validity.} AgentHarm labels harmful requests as unsafe, while WAInjectBench labels attack-derived text as unsafe even without explicit instructions. The EI/no-EI, threshold-sensitivity, and cross-model error analyses are exploratory. Our finite grid restricts allowance thresholds to at most 0.5 and blocking thresholds to at least 0.5. Other policies may yield different trade-offs.
\textbf{External validity.} Public benchmarks may overlap training data, and unsafe-input proportions affect precision, AP, and calibration. Results may differ for adaptive attacks, other languages, and multimodal inputs. Hosted model revisions and service behavior can change between runs. Protection also depends on the agent, environment, guard placement, and handling of blocked requests, requiring validation in the target system.

\section{Conclusion}
\label{sec:conclusion}
Our evaluation shows that aggregate rankings and calibration metrics can conceal attacks confidently classified as safe, and that strong performance on one security task does not consistently transfer to another. Policies selected under strict error limits allow few test inputs automatically, while choosing the allow and block thresholds separately increases automation mainly by rejecting more inputs. Developers should therefore select models for each security task, verify policy error limits on independent labeled data, and test which missed unsafe inputs reach a second judge, measuring its misses and benign false alarms on the reviewed inputs.

\section*{Acknowledgments}
OpenAI Codex assisted with study design and code writing.

\label{sec:body-end}
\section*{Data Availability}
The artifact provides experiment code, final predictions, benchmark partitions, configurations, and reproducible analyses (\url{https://github.com/yxsec/system-one-security-eval}).

\bibliographystyle{ACM-Reference-Format}
\bibliography{bibliography,references_extra}
\end{document}